\documentclass[aps,twocolumn,secnumarabic,superscriptaddress,balancelastpage,amsmath,amssymb,nofootinbib,titlepage]{revtex4-2}
\usepackage[english]{babel}
\usepackage[utf8]{inputenc}
\usepackage{amsmath}
\usepackage{graphicx}
\usepackage[colorinlistoftodos]{todonotes}
\usepackage{bibentry}
\usepackage{url}
\usepackage{rotating}
\usepackage{float}
\usepackage{color}
\usepackage[sort&compress]{natbib}
\usepackage[colorlinks=true,citecolor=blue]{hyperref}
\usepackage[caption=false]{subfig}
\usepackage{empheq}
\usepackage{bm}
\usepackage[normalem]{ulem}
\usepackage{siunitx}

\newcommand{\be}{\begin{equation}}
\newcommand{\ee}{\end{equation}}
\newcommand{\bea}{\begin{eqnarray}}
\newcommand{\eea}{\end{eqnarray}}
\newcommand{\bse}{\begin{subequations}}
\newcommand{\ese}{\end{subequations}}

\begin{document}
\title{Entanglement in an expanding toroidal Bose-Einstein condensate}

\author{Anshuman Bhardwaj} 
\affiliation{Department of Physics and Astronomy, Louisiana State University, Baton Rouge, LA 70803 USA}

\author{Ivan Agullo}
\affiliation{Department of Physics and Astronomy, Louisiana State University, Baton Rouge, LA 70803 USA}
\affiliation{Perimeter Institute for Theoretical Physics, 31 Caroline Street North, Waterloo, Ontario, Canada N2L 2Y5}
\author{Dimitrios Kranas}
\affiliation{Department of Physics and Astronomy, Louisiana State University, Baton Rouge, LA 70803 USA}
\author{Justin H. Wilson} 
\affiliation{Department of Physics and Astronomy, Louisiana State University, Baton Rouge, LA 70803 USA}
\author{Daniel E. Sheehy}
\affiliation{Department of Physics and Astronomy, Louisiana State University, Baton Rouge, LA 70803 USA}

\date{July 14, 2023}

\begin{abstract}
Recent experiments have employed rapidly expanding toroidal Bose-Einstein condensates (BECs) to mimic the inflationary  expansion in the early universe. One expected signature of the expansion in such experiments is spontaneous particle creation (of phonons) which is observable in density-density correlations. We study entanglement of these particles, which are known to result in a two-mode squeezed state. Using techniques for Gaussian states of continuous variable systems, we quantify the entanglement generated in this system, including effects such as decoherence and the use of
an initially squeezed state, which can suppress and enhance entanglement, respectively.  
We also describe a protocol to experimentally measure the correlations entering the covariance matrix, allowing an experimental quantification of the entanglement properties of the inflationary BEC.    
\end{abstract}

\maketitle 

\section{Introduction}

In 1966, L.~Parker made an important discovery~\cite{Parker:1966,Parker:1968mv}: The expansion of the universe can create particles out of the vacuum (see \cite{Schrodinger:1939} for earlier intuition about this phenomenon by E. Schr\"odinger). Parker considered Friedman-Lema\^itre-Robertson-Walker (FLRW) spacetimes that are asymptotically Minkowskian in the future and past, and showed that (non-conformal invariant) quantum fields that at early times are prepared in the vacuum state, generically end up in an excited state. The underlying translational invariance of the gravitation field leads to momentum conservation for the quantum field, which in turn implies that particles are created in pairs, with wavenumbers $\vec k$ and $-\vec k$. 
Phrased in a modern language,  Parker showed that, for each pair $(\vec k,-\vec k)$, the initial vacuum evolves to a {\em two-mode squeezed vacuum}, with squeezing intensity and squeezing angle determined  by the expansion history of the universe. This means, in particular, that the two particles in each created pair are {\em entangled}. 

Parker's ideas were formulated using the theory of linearized quantum fields propagating on a fixed gravitational background where the quantized fields are {\em test fields} which do not disrupt or modify in any way the underling geometry. This  same formalism was later applied to black holes by S.~Hawking in the mid seventies, leading to the Hawking effect \cite{Hawking:1974rv,Hawking:1974sw}, and also to the paradigm of cosmic inflation in the early eighties \cite{Starobinsky:1980te,Guth:1980zm,Albrecht:1982wi,Hawking:1981fz,Linde:1981mu,Linde:1983gd,Liddle:2000cg}. These two phenomena constitute important predictions of quantum field theory in curved spacetimes. In particular, the latter 
provides a possible explanation for the origin of the density perturbations in the early universe, which seeded the matter distribution we observe today. This would imply that  the  structures in our universe have  emerged from a process of squeezing of the quantum vacuum. 
This raises the question of how to test the quantum origin of this possibility.
Concretely, efforts to answer this have focused both on quantifying the genuine quantum correlations of squeezing due to inflation (i.e., measures of quantum entanglement generated) and how to observe 
it~\cite{dePutter:2019xxv,Martin:2015qta,Martin:2017zxs,Campo:2005sv,Maldacena:2015bha,Kanno:2017dci,Choudhury:2016cso,Martin:2021qkg,Agullo:2022ttg,Hsiang:2021kgh,Martin-Martinez:2012chf,Fuentes:2010dt}.

Independently,  W.~Unruh in Ref.~\cite{Unruh:1980cg} proved that the physics of quantum fields propagating on non-trivial geometries can be simulated in the lab, laying the groundwork for the field of analog gravity \cite{Barcelo:2005fc}. Analog models offer a test bed to recreate Parker's phenomenon of particle creation in the lab in a controlled manner, and to confirm its key predictions. In this paper, we focus attention on one of the simplest analog systems leading to particle creation \`a la Parker: a toroidal Bose-Einstein condensate (BEC) whose radius is rapidly growing.  This system has been realized
experimentally~\cite{Eckel:2017uqx,Banik:2021xjn}, and used to simulate an effectively one-dimensional inflationary universe \cite{Llorente:2019rbs,Bhardwaj:2020ndh,Eckel:2020qee}.  
These experiments observed the cosmological redshift and the damping that the expansion induces on density perturbations in the fluid. This is a promising platform to directly observe particle pair creation. Recent theory \cite{Bhardwaj:2020ndh} has analyzed the spectrum of particles created in this system under a finite period of exponential expansion and computed the signatures that these particles produce on density-density correlation functions. 

The focus of this paper is on entanglement. As mentioned above, entanglement is the quantum hallmark of the process of two-mode squeezing behind pair-creation, and observing it would assist in identifying and distinguishing the physical origin of the observed correlations. 
This work builds on this theoretically by probing the observability of the pair-creation process.
The rest of the paper is organized as follows: in Section~\ref{Section 2}, we briefly review how the perturbations inside a rapidly expanding toroidal BEC setup lead to spontaneous creation of phonon pairs in modes $(n,-n)$ in two-mode squeezed states (see Ref.~\cite{Bhardwaj:2020ndh} for more details). In Section~\ref{Section 3}, we first quantify the entanglement generated in this process by using an entanglement measure (logarithmic negativity) well-adapted to the physical setup. Next, this quantification allows us to analyze the way entanglement is affected by the presence of thermal noise and losses, ubiquitous in real experiments. Using the tools put forward in \cite{Agullo:2021vwj,Brady:2022ffk}, we show that noise and losses {\em degrade} the entanglement in the final state, possibly  eliminating it entirely and rendering a classical final state.
Next, following the ideas in Refs.~\cite{Agullo:2021vwj} and \cite{Brady:2022ffk}, we propose a way of amplifying the entanglement generated in this scenario, possibly compensating for the aforementioned deleterious effects. This is done by considering {\em stimulated} particle-creation using appropriate initial states---stimulated in the standard sense of atomic physics; to be contrasted with the spontaneous effects arising when the input is merely vacuum fluctuations. In particular, we study the use of single-mode squeezed inputs as a way of stimulating additional creation of entangled pairs, increasing the observability of this effect. In Sec.~\ref{Section 4}, we propose an experimental protocol to reconstruct the final state and to measure the entanglement it contains between phonons in the mode pair $(n,-n)$.  Section~\ref{sec:concl} provides some concluding remarks.  In
Appendix~\ref{AppendixA}, we present a brief review of Gaussian states and entanglement measures, and in
Appendix~\ref{appendix: C2 scale factor} we discuss a $C^2$-smooth expansion protocol for the toroidal BEC.

\section{Particle Creation in Toroidal BEC}\label{Section 2}

In this section, we briefly review and expand upon the work done in \cite{Bhardwaj:2020ndh}, that analyzed properties of a toroidal BEC with a time dependent radius, and the phenomenon of pair-creation of phonons therein---the reader is referred to \cite{Bhardwaj:2020ndh} for details omitted in this section.

The system experimentally created in \cite{Eckel:2017uqx,Banik:2021xjn} is made of a BEC with toroidal shape and time-dependent radius $R(t)$. In the thin-ring limit, variations in the condensate in the radial direction can be neglected, and the problem  becomes effectively one-dimensional, parametrized by the angle  $\theta$.  The BEC is described in terms of a complex scalar field $\hat \Phi(\theta,t)$~\cite{Pethick & Smith}, which can be decomposed in terms of density $\hat n(\theta,t)$ and phase $\hat\phi(\theta,t)$:
\be \label{Madelung representation}
\hat \Phi(\theta,t)=\sqrt{\hat n(\theta,t)}\, e^{i\, \hat\phi(\theta,t)}\, . 
\ee
This is the Madelung representation. We are interested in {\em linear perturbations}  $(\hat \phi_1(\theta,t),\hat n_1(\theta,t))$ of the phase and density, respectively, over a background $(\phi_0(t), n_0(t))$ describing the average phase and density, such that 
\be \label{perturbations}
\hat\phi(\theta,t)=\phi_0(t)+\hat \phi_1(\theta,t), \  \hat n(\theta,t)= n_0(t)+\hat n_1(\theta,t)\, . 
\ee
The  canonical commutation relation of the condensate $\hat \Phi$ imply that $(\hat n_1, \hat \phi_1)$ form a canonical operator-pair. The quantization of the pair $(\hat n_1, \hat \phi_1)$ is standard  \cite{Bhardwaj:2020ndh}. We begin with an expansion of these operators in terms of annihilation and creation operators ($\hat{a}_n,\hat{a}_n^\dagger$):
\begin{subequations}
\label{eq:phianddensity}
\begin{eqnarray}\label{Mode Expansion for Initial Phase}
\hspace{-0.3cm}	\hat{\phi}_{1}(\theta,t) & = &  \sqrt{\frac{U}{2\pi\mathcal{V}_0\hbar}} \sum_{n=-\infty}^{\infty}\big[e^{i  n\theta}\chi_{n}(t)\, \hat{a}_{n}+{\rm h.c.}\big],~~~~ \\ \label{Mode Expansion for Initial Density}
\hspace{-0.3cm}	\hat{n}_{1}(\theta,t) & = &-\frac{\hbar\mathcal{V}(t)}{ U}\frac{d}{dt}\hat{\phi}_{1}(\theta,t) 
, 
\end{eqnarray}
\end{subequations}
where h.c.\ indicates Hermitian conjugate, $U=\frac{4\pi a_{s}\hbar^{2}}{M}$ is the interaction parameter,  with $a_{s}$ being the scattering length, $M$ is the mass of the atoms in the condensate, $2\pi \mathcal{V}(t)$ is the volume of the condensate, and $\mathcal{V}_0\equiv \mathcal{V}(0)$. The time-dependent functions $\chi_{n}(t)$  are the mode functions, and they form a basis of the vector space of complex solutions to the equations of motion~\cite{Bhardwaj:2020ndh}
\be\label{MukhanovSasakiQP}
	\ddot{\chi}_{n} + \big(1+\gamma\, \big)\frac{\dot{R}}{R}\dot{\chi}_{n} + \alpha\, \frac{n^{2}c^{2}}{R^{2}}\chi_{n} = 0.  \\
\ee
In these equations, that are analogous to the Mukhanov-Sasaki equations~\cite{Sasaki:1983kd,Kodama:1985bj,Mukhanov:1988jd} from cosmology, 
$\gamma$ and $\alpha$ are corrections due to quantum pressure,  which generally 
depend on the density, the radius of the ring and 
the mode index. Following \cite{Bhardwaj:2020ndh}, we will approximate $\gamma$ and $\alpha$ by constants and take 
$0<\gamma<1$ and 
$\alpha = 1$.
Let us write the time-dependent radius of the toroid as $R(t)=R_0\, a(t)$, with $R_0$ being a constant. Then, if the mode functions are chosen to be normalized such that $( \chi_{n}\dot  \chi^{*}_{n}- \chi^{*}_{n}\dot  \chi_{n}) = i\,  a(t)^{-(1+\gamma)}$ for all $n$ at any instant, then this normalization is preserved throughout the evolution. Furthermore, $\hat{a}_{n}$ and $\hat{a}^{\dagger}_{n}$ in \eqref{eq:phianddensity} satisfy the algebra of creation and annihilation operators, i.e., $[\hat{a}_{n},\hat{a}^{\dagger}_{n'}]=\delta_{nn'}$, and $[\hat{a}_{n},\hat{a}_{n'}]=0$.

Notably, Eqs.~\eqref{eq:phianddensity} and \eqref{MukhanovSasakiQP} are formally analogs to the equations describing the propagation of a scalar field in a spatially-flat FLRW universe, with scale factor $a(t)$.\footnote{If the quantum pressure $\gamma$ were to vanish, Eq.~\eqref{MukhanovSasakiQP} would be equivalent to the equation one would find for the modes of a massless, minimally coupled scalar field in a FLRW spacetime with one spatial dimension. But such a field is conformally invariant, and since the FLRW geometry is conformally flat, there would be no particle creation in that situation. The presence of  $\gamma\neq 0$ in \eqref{MukhanovSasakiQP} breaks conformal invariance and makes it possible for phonon pair-creation to occur.}  Hence, the system under consideration makes it possible to recreate the physics of quantum fields propagating in an expanding universe, by appropriately engineering a time-dependent radius $R(t)$ of the toroidal BEC, as was done in \cite{Eckel:2017uqx,Banik:2021xjn}. 

Following the experimental platform described in \cite{Eckel:2017uqx,Banik:2021xjn}, we consider $R(t)$ that is time-independent in the past, then varies monotonically, and finally becomes constant again. These early- and late-time  regions where $R(t)$ is time-independent, are the ``in'' and ``out'' regions, respectively, in which the system is stationary and there is a well-defined notion of ground state---the in and out vacuum, $|0_{\rm in}\rangle $ and  $|0_{\rm out}\rangle$, respectively---each associated with a set of annihilation and creation operators, which we will denote as $(\hat{a}^{{\rm (in)}},\hat{a}^{{\rm (in)}\dagger})$ and $(\hat{a}^{{\rm (out)}},\hat{a}^{{\rm (out)}\dagger})$, respectively.

\begin{figure}[h!]
	\begin{center}
	\includegraphics[width=1.0\columnwidth]{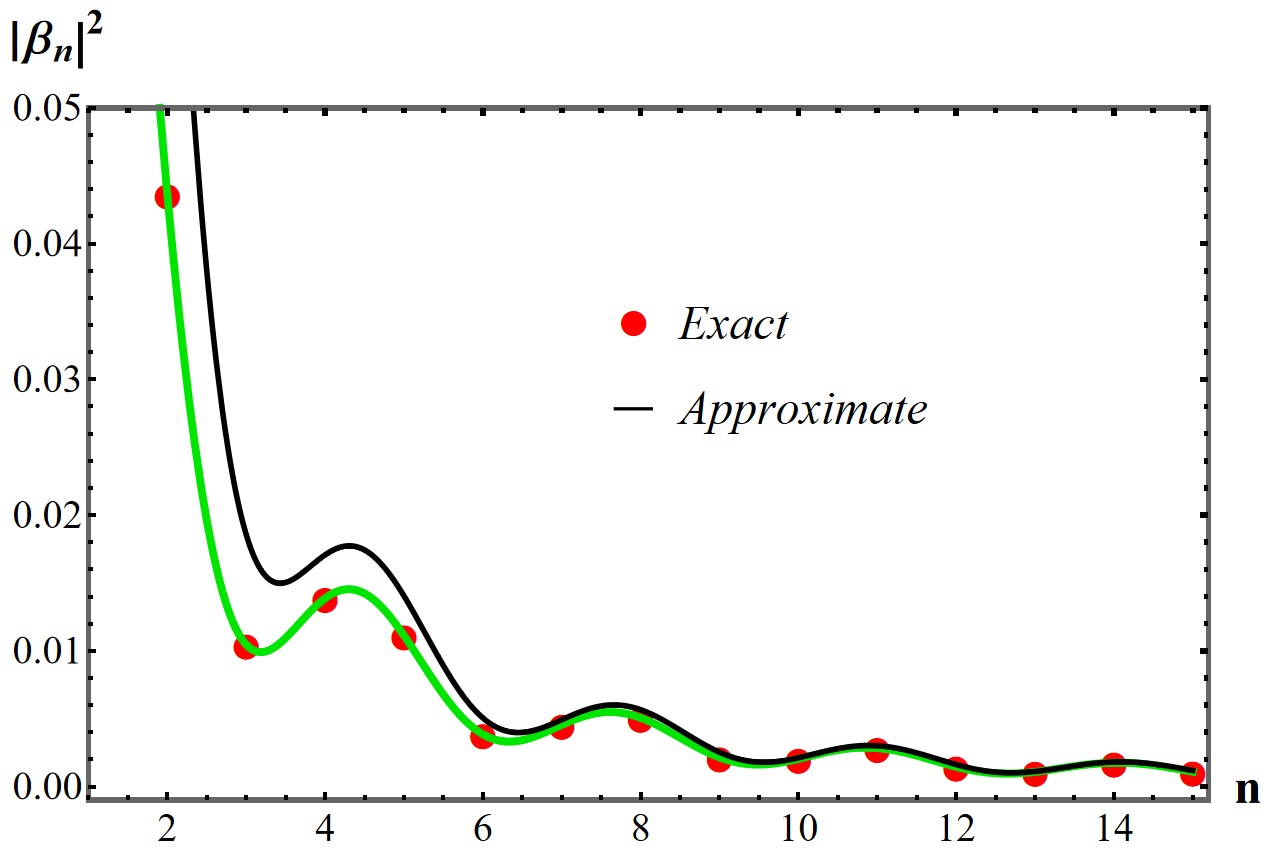}
	\end{center}
	\caption{(Color Online) Plot of the mean number of quanta $|\beta_{n}|^{2}$ created in mode $n$. The red circles indicate the exact thoretical result from the absolute value squared of Eq.~(\ref{beta_n}), and the black curve shows the approximate result in Eq.~(\ref{Asymptotic Particle Creation}), valid in the limit of small $\gamma$ and large $n$. The green curve shows $|\beta_{n}|^{2}$ as a continuous function to emphasize its overall dependence on the discrete mode indices. For this plot, we took the quantum pressure to be $\gamma=0.5$, the speed of sound to be $c = \qty{2}{\mm/\s}$, the initial radius to be $R_{0} = \qty{10}{\um}$, the expansion timescale to be $\tau = \qty{6.21}{\ms}$ and the duration of expansion to be $t_{f} = \qty{10}{\ms}$.}
	\label{fig1}
\end{figure}
 
One question of interest is the following: if the condensate is prepared in the mean-field ground state at early times, what is the state at late times? The answer to this question is well-known \cite{Parker:1966,Parker:1968mv,Bhardwaj:2020ndh}  
 \begin{equation}\label{SqueezedParticleCreation}
\hat U |0_{\rm in}\rangle = \prod_{n\in\mathcal{Z}}^{}\frac{1}{\sqrt{|\alpha_{n}|}}e^{\frac{\beta^*_{n}}{2\alpha^*_{n}}\hat{a}^{{\rm (out)}\dagger}_{n}\hat{a}^{{\rm (out)}\dagger}_{-n}} |0_{\rm out}\rangle,
\end{equation}
where $\hat U $ is the time-evolution operator, and $\alpha_{n}$ and $\beta_{n}$ are the Bogoliubov coefficients relating the in and out creation and annihilation operators
 \begin{equation}
\label{eq:bogTransf}
 \hat{a}^{{\rm (out)}}_{n}=\alpha_{n}\, \hat{a}^{{\rm (in)}}_{n} + \beta^{*}_{n}\, \hat{a}^{{\rm (in)}\dagger}_{-n} \, .
\end{equation}
By expanding the exponential in \eqref{SqueezedParticleCreation}, we see that the result of evolving the in vacuum produces a state at late times that is made of linear combinations of states of the form $(\hat{a}^{{\rm (out)}\dagger}_{n}\hat{a}^{{\rm (out)}\dagger}_{-n})^N |0_{\rm out}\rangle$. These states contain $2N$ phonons, half of them with label $n$ and the other half with label $-n$. This implies that phonons are created in pairs $(n,-n)$. Since modes with label $n$ and $-n$ describe  counter- and clockwise propagating plane waves, respectively, the creation of these pairs respects angular momentum conservation. Furthermore, it is not difficult to check that there are $|\beta_n|^2$  phonons, on average, in the mode $n$ at late times. In the next section, we will write the final state in a different form, which will make manifest that \eqref{SqueezedParticleCreation} is a Gaussian state resulting  by applying a process of two-mode squeezing to the vacuum for each pair $(n,-n)$, and we will describe a way of quantifying the entanglement in this state.

Our next task is to choose a specific function for
the toroidal BEC radius $R(t)$ entering Eq.~(\ref{MukhanovSasakiQP}). 
Following Ref.~{\cite{Bhardwaj:2020ndh}}, we choose:
\begin{equation}
R(t) = \begin{cases} R_{0}, & \text{for $t<0$},
\cr
R_{0}e^{t/\tau}, & \text{for $0<t<t_{f}$},
\cr
R_{0}e^{t_{f}/\tau}, & \text{for $t>t_{f}$}.
\end{cases}
\label{Eq:expansionchoice}
\end{equation}
Thus, for early times $t<0$, the radius is a constant $R_{0}$, with the
exponential expansion beginning at $t=0$.  In this exponential regime, we can write
 $R(t)=R_{0}\, a(t)$ with $a(t)= e^{t/\tau}$ being the scale factor, thereby
 mimicking an inflationary universe, where the constant $\tau$ plays the role of the ``Hubble time''. For later times $t>t_{f}$, the radius is again a constant $R_{f}=R_{0}e^{t_{\text{f}}/\tau}$.

 It is important to emphasize that the conceptual analysis and the tools presented in this article are applicable to any form of the expansion history $R(t)$. We use the simple expression \eqref{Eq:expansionchoice} merely for illustrative purposes. All the plots shown below can be straightforwardly re-computed for $R(t)$ adapted to concrete experimental platforms.

The simplicity of this choice leads to an issue: 
the function $R(t)$ in Eq.~(\ref{Eq:expansionchoice}) is continuous but not differentiable at $t=0$ and $t=t_f$.  
We expect that, in a real experiment, the ramp-up and down will be smoother (with its own time scale), and as a result, this abrupt model expansion we include here may lead to spurious particle production outside of the central inflationary regime.
This issue was pointed out by Glenz and Parker (GP), who argued that such artifacts can be avoided as long as $R(t)$ is at least $C^2$~\cite{Glenz:2009zn}.
The experiments of
Refs.~\cite{Eckel:2017uqx,Banik:2021xjn} use an error-function profile for the trap
expansion to minimize such slope discontinuities.  

To justify our use of Eq.~(\ref{Eq:expansionchoice}) despite this issue, we note that, as emphasized by GP, artifacts of a $C^{0}$ scale factor 
are most significant at large mode index, i.e., density fluctuations at short distances.  We argue that, since such high mode indices are probably inaccessible to real toroidal BEC experiments, Eq.~(\ref{Eq:expansionchoice}) is sufficient for our goal of
studying entanglement of particle production.  In addition, Eq.~(\ref{Eq:expansionchoice}) has the advantage of providing closed analytic
expressions for the Bogoliubov coefficients $\alpha_{n}$ and
$\beta_{n}$~\cite{Bhardwaj:2020ndh}.  

Nevertheless, for completeness, in Appendix~\ref{appendix: C2 scale factor} we consider an expansion history $R(t)$ of differentiability class $C^{2}$, also containing an inflationary period.  
There, we show that the resulting  mean number of created quanta $|\beta_n|^{2}$  behaves universally as $n^{-\gamma}$ for  sufficiently large $n$, directly reflecting the dissipative role of the
quantum pressure $\gamma$ in the equation of motion (\ref{MukhanovSasakiQP}).

Now returning to our model Eq.~(\ref{Eq:expansionchoice}) for the BEC
radius, we quote the resulting Bogoliubov 
coefficients~\cite{Bhardwaj:2020ndh}:
\begin{widetext}
\begin{eqnarray}\label{alpha_n}
\alpha_{n} & = & \frac{1}{2}\frac{e^{-\frac{t_{f}}{2\tau}}}{J_{\frac{1+\gamma}{2}}(z_{0})J_{\frac{1-\gamma}{2}}(z_{0})+J_{-\frac{1+\gamma}{2}}(z_{0})J_{\frac{-1+\gamma}{2}}(z_{0})} \nonumber \\
& \times & \bigg[ \Big\{J_{\frac{1-\gamma}{2}}(z_{0})+iJ_{-\frac{1+\gamma}{2}}(z_{0})\Big\}\Big\{J_{\frac{1+\gamma}{2}}(z_{f}){-}iJ_{\frac{-1+\gamma}{2}}(z_{f})\Big\} + \Big\{J_{\frac{-1+\gamma}{2}}(z_{0})-iJ_{\frac{1+\gamma}{2}}(z_{0})\Big\}\Big\{J_{-\frac{1+\gamma}{2}}(z_{f}){+}iJ_{\frac{1-\gamma}{2}}(z_{f})\Big\} \bigg],~~~~~ \\ \label{beta_n}
\beta_{n} & = & \frac{1}{2}\frac{e^{-\frac{t_{f}}{2\tau}}}{J_{\frac{1+\gamma}{2}}(z_{0})J_{\frac{1-\gamma}{2}}(z_{0})+J_{-\frac{1+\gamma}{2}}(z_{0})J_{\frac{-1+\gamma}{2}}(z_{0})} \nonumber \\
& \times & \bigg[ \Big\{J_{\frac{1-\gamma}{2}}(z_{0})+iJ_{-\frac{1+\gamma}{2}}(z_{0})\Big\}\Big\{J_{\frac{1+\gamma}{2}}(z_{f}){+}iJ_{\frac{-1+\gamma}{2}}(z_{f})\Big\} + \Big\{J_{\frac{-1+\gamma}{2}}(z_{0})-iJ_{\frac{1+\gamma}{2}}(z_{0})\Big\}\Big\{J_{-\frac{1+\gamma}{2}}(z_{f}){-}iJ_{\frac{1-\gamma}{2}}(z_{f})\Big\} \bigg],~~~~~
\end{eqnarray}
\end{widetext}
where $J_{n}(x)$ are Bessel functions of the first kind, and we define parameters $z_{0}\equiv\frac{|n|c\tau}{R_{0}}$ and $z_{f}\equiv\frac{|n|c\tau}{R_{f}}$. 

Next, we examine various limits for the particle production $|\beta_n|^{2}$.  
A simple approximation for $|\beta_n|^{2}$ (valid in the limit $\gamma\ll1$ and $n\gg1$), is given  by {\cite{Bhardwaj:2020ndh}} 
\begin{equation}\label{Asymptotic Particle Creation}
|\beta_{n}|^{2}\approx\frac{1}{n^2}\, \Big(\frac{\gamma}{4}\Big)^{2}\Big(\frac{c\tau}{R_{0}}\Big)^{-2}\big[1+a_f^{2}-2a_f\cos\big(2n\theta_{\rm H}\big)\big], 
\end{equation}
where $a_f=e^{\frac{t_{f}}{\tau}}$ is the ratio of final and initial radii $R_{f}/R_{0}$, and $\theta_{\rm H}=\frac{c\tau}{R_{0}}\big(1-a_f^{-1}\big)$ is the angular horizon size at the end of the expansion. Although this approximation is  obtained as a large $n$ limit, it  is quite accurate for $n \gtrsim 3$ in relevant parameter ranges as can be seen in Fig.~\ref{fig1}. This approximation makes it clear that oscillations in the particle creation  number shown in Fig.~\ref{fig1} reflect the value of the angular size of the Hubble horizon $\theta_{\rm H}$ (see Ref.~\cite{Bhardwaj:2020ndh} for details). Note also that the pair creation in the expanding ring does not produce a particle distribution with a black-body spectrum, as happens for analog models for which causal horizons are present. 

We can still recover the inflationary regime $|\beta_n|^2 \sim n^{-\gamma}$ with assumptions which simplify the Bessel functions with arguments $z_0$ and $z_f$ in Eq.~(\ref{beta_n}).
 First, we assume that the modes $n$ are initially well within the horizon. In this limit, $z_{0}\to\infty$, and the Bessel functions with argument $z_0$ 
  become: $J_{\alpha}(z_{0})\sim\sqrt{\frac{2}{\pi z_{0}}}\cos(z_{0}-\frac{\alpha\pi}{2}-\frac{\pi}{4})$. Second,  if we assume that the toroid expands for a very long time (an experimentally challenging assumption)
the modes $n$ exit the horizon and attain very large wavelengths compared to the horizon size, i.e., $R_{f}/n\gg c\tau$, and the Bessel functions with argument $z_f$
become: $J_{\alpha}(z_{f})\sim\frac{1}{\Gamma(\alpha+1)}(\frac{z_{f}}{2})^{\alpha}$. In this regime, the particle creation   number in Eq.~(\ref{beta_n}) takes the following form:
\begin{equation}\label{Asymptotic Particle Creation Superhorizon}
|\beta_{n}|^{2} \approx \frac{\pi}{4}\Big(\frac{R_{f}}{2c\tau}\bigg)^{\gamma}\frac{\cos^{2}(z_{0}-\frac{\pi\gamma}{4})+\cos^{2}(z_{0}+\frac{\pi\gamma}{4})}{\sin^{2}\big(\frac{1+\gamma}{2}\pi\big)\Gamma^{2}\big(\frac{1+\gamma}{2}\big)}\, n^{-\gamma}.
\end{equation}

\begin{figure}[h!]
	\begin{center}
	\includegraphics[width=1.0\columnwidth]{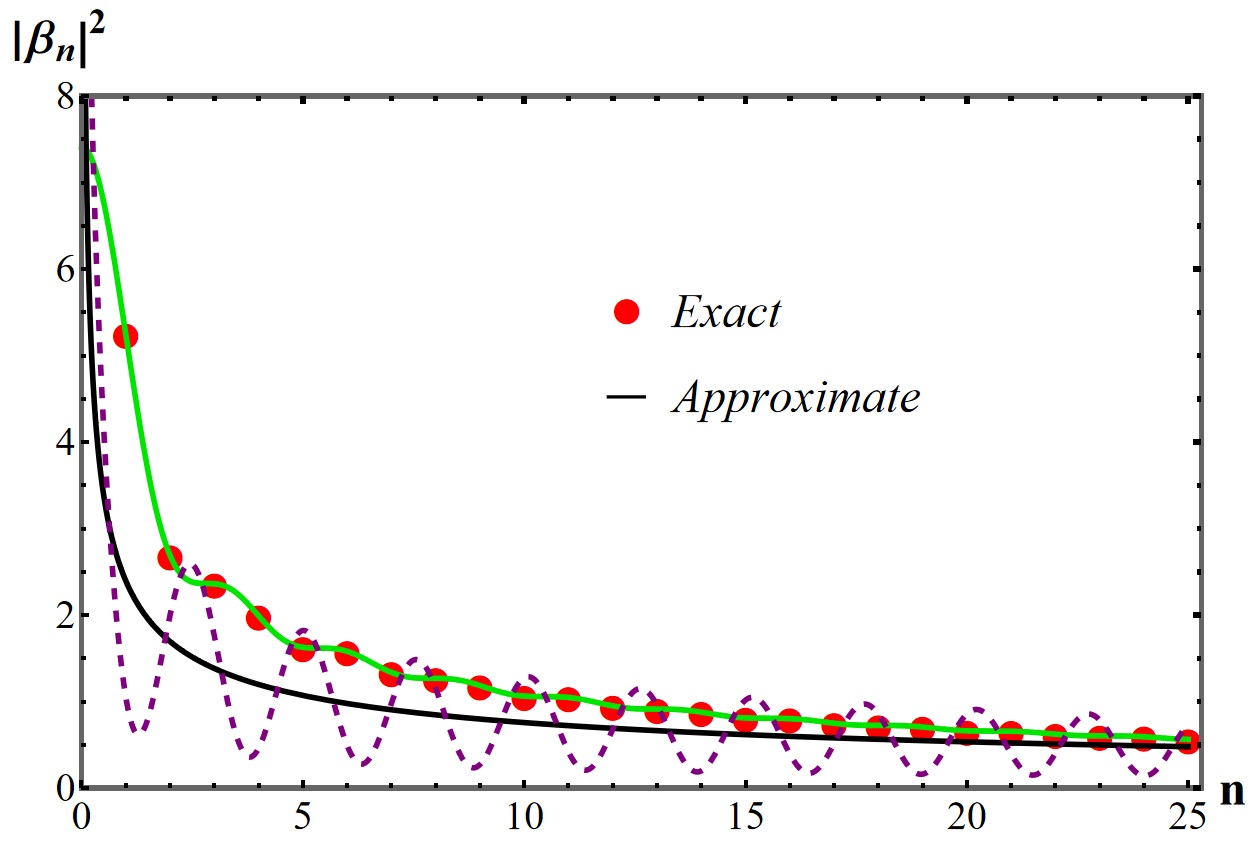}
	\end{center}
	\caption{(Color Online) Plot of the  mean number of quanta $|\beta_{n}|^{2}$ created in mode $n$. The red circles indicate the exact theoretical result 
 from the absolute value squared of Eq.~(\ref{beta_n}), and the green curve is its continuous version to emphasize its overall dependence on the discrete mode indices. We compare this with the dashed purple curve that shows the approximate result in Eq.~(\ref{Asymptotic Particle Creation Superhorizon}), where the black curve is obtained by averaging the two cosine square terms in (\ref{Asymptotic Particle Creation Superhorizon}). For this plot, we took the same parameters as Fig.~\ref{fig1}, but with a much longer expansion time $t_{f} = \qty{1e4}{\ms}$.}
	\label{fig1 alt}
\end{figure}

As can be seen from Fig.~\ref{fig1 alt}, this approximation fits well with  the exact expression for $|\beta_{n}|^{2}$ for large expansion times, leading to the universal behavior of $n^{-\gamma}$ as predicted by GP \cite{Glenz:2009zn}. However, for smaller expansion times (where
$R_f$ is not so large), 
it is more appropriate 
to use the $n^{-2}$ asymptotic form in Eq.~\eqref{Asymptotic Particle Creation}. In addition to being experimentally  challenging, long expansions lead to an increase in the coherence length\footnote{For a BEC, the coherence length is defined as the length scale over which the condensate maintains coherence in its density given by $\xi=\frac{\hbar}{\sqrt{2}Mc}$, where $M$ is the mass of the bosonic atoms and $c$ is the speed of sound.}
and a decrease of the ring width with increasing time, to the point that the hydrodynamic limit (which requires the mode wavelength to be large 
compared to the coherence length) is violated.
Intuitively, long expansion times decrease the density to a point where individual atoms are spread out sufficiently that they cannot interact and form the coherent mean-field ground state.

\begin{figure}[h!]
	\begin{center}
	\includegraphics[width=\columnwidth]{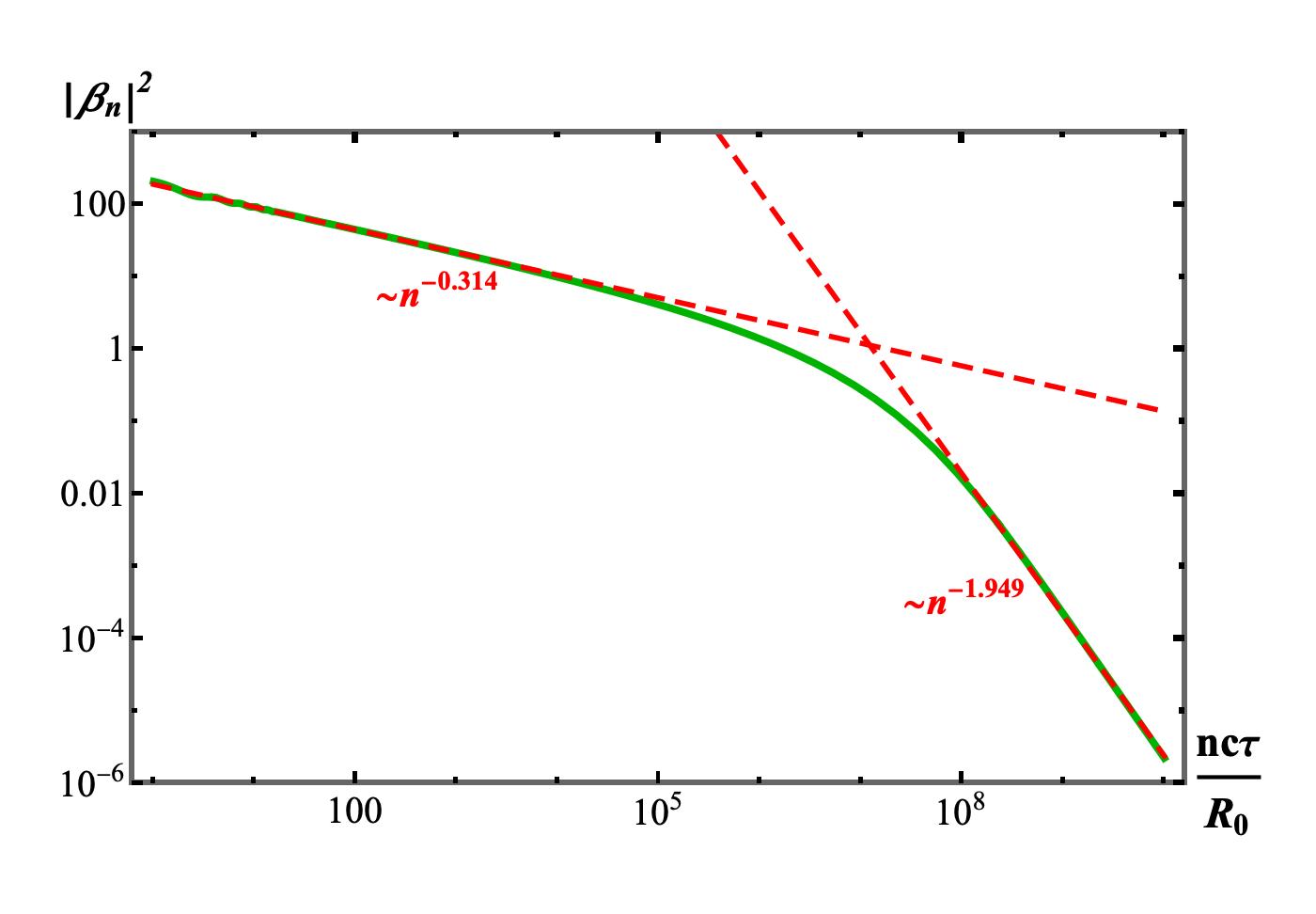}
	\end{center}
	\caption{(Color Online) Log-log plot of the  mean number of quanta $|\beta_{n}|^{2}$ created in mode $n$,
 for the case of ${\rm e}^{t_f/\tau} = 10^8$ 
 and $\gamma = 0.3$ (solid green curve).  
 The red dashed lines are linear
 fits with the fitted power law shown on the plot, showing that $|\beta_{n}|^{2}$ indeed approximately exhibits 
 $n^{-2}$ behavior at large $n$ and 
 $n^{-\gamma}$ behavior at smaller $n$, consistent with 
 Eqs.~(\ref{Asymptotic Particle Creation})
 and (\ref{Asymptotic Particle Creation Superhorizon}),
 respectively.
 }
	\label{fig3new}
\end{figure}

The preceding two regimes of the particle creation  number, 
Eqs.~(\ref{Asymptotic Particle Creation}) and (\ref{Asymptotic Particle Creation Superhorizon}), 
approximately hold in the regimes $n\gg \frac{R_f}{c\tau}$ and 
$\frac{R_0}{c\tau}\ll n\ll \frac{R_f}{c\tau}$, respectively. The crossover between these two
regimes can be clearly seen in a log-log
plot of $|\beta_n|^2$ vs. $n$, as shown in 
Fig.~\ref{fig3new}.
In the third regime 
of $n \ll \frac{R_0}{c\tau}$, $|\beta_{n}|^{2}$ is approximately independent of 
mode index. 
In Ref.~\cite{Bhardwaj:2020ndh}, it was shown that the  pair-creation 
represented by $|\beta_n|^2$  produces  distinctive  features in density-density correlations of the condensate, which could be observed in the laboratory. The main effect is a kink-like feature  in the density-density two-point correlation function, located at an angular separation determined by the angular size of the horizon $\theta_{\rm H}$. The amplitude of this kink increases with 
increasing quantum pressure and BEC temperature. In the zero temperature limit, the kink degenerates into a ``cusp''-like feature.

Having described both the perturbations in an expanding toroidal BEC and how such expansion leads to spontaneous phonon pair creation in a two-mode squeezed state (\ref{SqueezedParticleCreation}) determined by $\beta_{n}$ given in Eq.~(\ref{beta_n}), we are ready to discuss the entanglement of these phonon pairs, i.e., how to quantify it and what factors lead to its degradation and enhancement.

\section{Entanglement}\label{sec:entanglement}\label{Section 3}

In this section, we reformulate the evolution of phonon perturbations propagating in the expanding toroidal BEC, using the language of  continuous variable quantum systems and Gaussian states~\cite{Serafini:2017}  (Ref.~\onlinecite{Brady:2022ffk} for applications to analog gravity). This formalism is useful because the equations of motion \eqref{MukhanovSasakiQP} of phonon perturbations are linear, and therefore preserve the Gaussianity of the quantum states.  

Using this language, the evolution of phonon perturbations is  described by a collection of two-mode squeezers, each acting on a pair $(n,-n)$ of modes. Classically, a two-mode squeezer represents a non-energy conserving process, able to amplify (or damp) waves. Quantum mechanics adds two key features. On the one hand, even the vacuum can be amplified. On the other hand, the amplification process creates pair-wise entanglement between quanta. The formalism used in this section provides an efficient toolbox to quantify this entanglement.

Gaussian states include vacua, coherent, thermal, and squeezed states. Therefore, although our analysis is restricted, the family of Gaussian states is sufficiently general to approximately 
describe most of the states one can create and manipulate in the laboratory.

\subsection{Evolution: two-mode squeezers and pair creation}

The background condensate (described by $n_0$ and $\phi_0$) is homogenous, in the sense that neither $n_0$ nor $\phi_0$ depend on the location $\theta$ along the ring. Consequently, the evolution of 
condensate perturbations $\chi(\theta,t)$ is such that Fourier modes with spatial dependence $e^{-i n \theta}$, $n\in \mathbb{Z}$, evolve independently of each other. This is manifest in Eq.~\eqref{MukhanovSasakiQP}.  For a fixed $n$, there are only two modes with such spatial dependence, namely $\chi_n$ and $\chi^*_{-n}$, using the notation introduced in the previous section. Hence, the evolution of the perturbations factors out in the evolution of pairs $(n,-n)$ of modes, with no interaction among different pairs. Since each pair corresponds to a quantum mechanical linear system with two degrees of freedom, we can apply the formalism summarized in Appendix~\ref{appendix: Gaussian states} (we refer the reader to this Appendix for details omitted here). 

Let  $\hat{{\bf A}}^{\rm (in)}_n$ be the (column) vector of in creation and annihilation operators for a  pair $(n,-n)$ of modes 
 \be \hat{{\bf A}}^{\rm (in)}_{n}\equiv (\hat{a}^{\rm (in)}_{n},\hat{a}^{\rm (in)\dagger}_n, \hat{a}^{\rm (in)}_{-n},\hat{a}^{\rm (in)\, \dagger}_{-n})^{\top}\, , \ee
and let  $\hat{{\bf A}}^{\rm (out)}_n$ be similarly defined for the out modes.  Expression \eqref{eq:bogTransf} above implies that the  matrix ${\bf S}_{{\rm ({\bf A}},n)}$ describing the in-out scattering process, 
$\hat{{\bf A}}^{\rm (out)}_{n}={\bf S}_{{\rm ({\bf A}},n)}\cdot \hat{{\bf A}}^{\rm (in)}_{n}$, is
\be 
{\bf S}_{{\rm ({\bf A}},n)}= \begin{bmatrix}
	\alpha_{n} & 0 & 0 & \beta_{n}^* \\
	0 & \alpha^*_{n} & \beta_{n} & 0 \\
	0 & \beta_{n}^* & \alpha_{n} & 0 \\
	\beta_{n} & 0 & 0 & \alpha^{*}_{n} 
\end{bmatrix}. 
\ee
This is precisely the form of a {\em two-mode squeezer} (see, e.g., Appendix A of \cite{Brady:2022ffk} for a short summary of this type of symplectic transformation). From ${\bf S}_{{\rm ({\bf A}},n)}$, we can obtain the scattering matrix $ {\bf S}_n$ describing the evolution of canonical pairs 
\begin{subequations}
\bea
\hat{x}_n&=&\frac{1}{2}(\hat{a}_{n}+\hat{a}^{\dagger}_{n}),
\\
\hat{p}_n&=&-\frac{i}{2}(\hat{a}_{n}-\hat{a}^{\dagger}_{n}),
\eea
\label{eq:connection}
\end{subequations}
by simple multiplication with the ``change of basis matrix'' $\bf B$, ${\bf S}_n={\bf B}\cdot {\bf S}_{{\rm ({\bf A}},n)}\cdot (\bf B)^{-1}$, where $\bf B$ is written in Eq.~\eqref{B}
of Appendix \ref{AppendixA}.

In experimental settings, the initial state of the system is well approximated by a thermal state in equilibrium with the environment at temperature $T$. This is a mixed Gaussian state, which means it
is completely determined by its first (${\bm \mu}$) and second (${\bm \sigma}$) moments, with
the latter being termed the covariance matrix.  (See Appendix~\ref{appendix: Gaussian states}, where
these are defined in Eqs.~(\ref{Eq:mudef}) and (\ref{Eq:sigmadef}), respectively.)  In the 
present case, we find:
\bea
{\bm \mu}_{(n)}^{\rm (in)}&=&(0,0,0,0)^{\top}\, , \\
{\bm \sigma}_{(n)}^{\rm (in)}&=&(1+2 n_B(E_n)) \, \mathbb{I}_4 \, ,\eea
where $\mathbb{I}_4$ is the identity matrix and  $n_{\text{B}}(x)=(e^{\beta x}-1)^{-1}$ is the mean number of thermal quanta, given by the Bose-Einstein distribution with 
$\beta=(k_{\text{B}}T)^{-1}$, and $k_{\text{B}}$ the 
Boltzmann constant.  Here, the Bogoliubov
energy dispersion is 
$E_{n}=\sqrt{\epsilon_{n}(\epsilon_{n}+2Mc^{2})}$,
where $c$ is the speed of sound and $\epsilon_{n}=\frac{\hbar^{2}n^{2}}{2MR^{2}}$ is the single particle energy in the toroid.  In the
hydrodynamic limit of interest here, 
in which the mode wavelength is large compared
to the coherence length,
we approximate $E_n \simeq \frac{\hbar c}{R}|n|$.
From this, we obtain the final state, which is again Gaussian and described  by 

\begin{widetext} 
\bea {\bm \mu}_{(n)}^{\rm (out)}&=&{\bf S}_n\cdot {\bm \mu}_{(n)}^{\rm (in)}=(0,0,0,0)^{\top}\, \label{outstate} \\
{\bm \sigma}_{(n)}^{\rm (out)}&=&{\bf S}_n\cdot{\bm \sigma}_{(n)}^{\rm (in)}\cdot{\bf S}_n^{\top} = (1+2n_{\text{B}}) \begin{bmatrix}
|\alpha_{n}|^{2}+|\beta_{n}|^{2} & 0 & \alpha_{n}\beta_{n}+\alpha^{*}_{n}\beta^{*}_{n} & -i(\alpha_{n}\beta_{n}-\alpha^{*}_{n}\beta^{*}_{n}) \\
0 & |\alpha_{n}|^{2}+|\beta_{n}|^{2} & -i(\alpha_{n}\beta_{n}-\alpha^{*}_{n}\beta^{*}_{n}) & -(\alpha_{n}\beta_{n}+\alpha^{*}_{n}\beta^{*}_{n}) \\
\alpha_{n}\beta_{n}+\alpha^{*}_{n}\beta^{*}_{n} & -i(\alpha_{n}\beta_{n}-\alpha^{*}_{n}\beta^{*}_{n}) & |\alpha_{n}|^{2}+|\beta_{n}|^{2} & 0 \\
-i(\alpha_{n}\beta_{n}-\alpha^{*}_{n}\beta^{*}_{n}) & -(\alpha_{n}\beta_{n}+\alpha^{*}_{n}\beta^{*}_{n}) & 0 & |\alpha_{n}|^{2}+|\beta_{n}|^{2}  
\end{bmatrix} \, . \nonumber
\eea
\end{widetext}
Note the Bogoluibov coefficients $\alpha_n$ and $\beta_n$ were written in Eqs.~(\ref{alpha_n}) and (\ref{beta_n}) above for the dynamical expansion given by Eq.~(\ref{Eq:expansionchoice}). Here and below we have
suppressed the argument of the
Bose-Einstein distribution, which is always $E_n$.
This covariance matrix is of the form
\be {\bm \sigma}_{(n)}^{\rm (out)}=\begin{bmatrix} {\bm\sigma}^{\rm (red)}_{(n)} &{\bf  C}_{(n)} \\{\bf C}_{(n)} ^{\top} & {\bm\sigma}^{\rm (red)}_{(-n)}  \end{bmatrix}  \, , 
\ee
where  ${\bm\sigma}^{\rm (red)}_{(n)}$ and ${\bm\sigma}^{\rm (red)}_{(-n)}$ are the covariance matrices of the reduced state describing the
modes $n$ and $-n$, respectively, which are equal to each other. The matrix  ${\bf C}_{(n)}$ encodes the correlations between these two modes. We will see in the next section that these correlations contain entanglement for sufficiently low environment temperatures. 

From Eq.~\eqref{outstate}, we extract all predictions about the final state such as  the mean number of phonons in one of the two modes, say the mode $n$: 
\be
  \langle \hat{N}_{n} \rangle=\frac{1}{4}\text{Tr}\{\bm\sigma^{\rm (red)}_{(n)}\}+\frac{1}{2}(\bm{\mu}^{\rm (red)}_{(n)})^\top\bm{\mu}^{\rm (red)}_{(n)} -\frac{1}{2},\label{eq:mean_quanta}
\ee
where 
\bea
\bm{\mu}^{\rm (red)}_{(n)}&=&(0,0)\, , \nonumber \\
{\bm\sigma}^{\rm (red)}_{(n)}&=&(1+2n_{\text{B}})\, (|\alpha_{n}|^{2}+|\beta_{n}|^{2}) \, \mathbb{I}_2 , 
\eea
are the first moments and covariance matrix of the reduced state describing the mode $n$ alone. 
We obtain
\bea
 \langle \hat{N}_{n} \rangle&=&\frac{1}{2} \big[ (1+2n_{\text{B}})\,  (|\alpha_{n}|^{2}+|\beta_{n}|^{2})-1\big], \\&=&n_{\text{B}} +|\beta_{n}|^{2} +2 n_{\text{B}}|\beta_{n}|^{2}\, ,  \eea
where in the second line we have used the identity $|\alpha_{n}|^{2}-|\beta_{n}|^{2}=1$. This last expression offers a simple interpretation. On the one hand, we see that, in the zero temperature limit, $n_{\text{B}}\to 0$, we obtain $\langle \hat{N}_{n} \rangle=|\beta_{n}|^{2}$, as expected. For finite temperature, $ \langle \hat{N}_{|n|} \rangle$ has three contributions. The first one, given  by $n_{\text{B}}$,   simply corresponds to the thermal quanta already present in the initial state. The second term,  $|\beta_{n}|^{2}$, 
corresponds to the quanta created from the vacuum. The third term contains the product $ n_{\text{B}}|\beta_{n}|^{2}$,  which has the interpretation of {\em stimulated or induced phonon creation} (i.e., the mere presence of initial quanta induces further pair-creation). 

The number of quanta in the mode $-n$ has exactly the same value, in such a way that $\langle \hat{N}_{n} \rangle- \langle \hat{N}_{-n} \rangle$ remains constant in the course of time. In other words, quanta are created in pairs $(n,-n)$. 

\subsection{Entanglement}

In this subsection, we  investigate under what conditions the out state written in Eq.~(\ref{outstate}) for the $(n,-n)$ pair  is entangled. Appendix~\ref{appendix: Gaussian states} contains a summary of a few ways of answering this question (see \cite{Serafini:2017} for further details on entanglement measures for Gaussian states; and see also \cite{Agullo:2021vwj,Brady:2022ffk} for applications to Hawking radiation). 
 
The state (\ref{outstate}) is a mixed state for any non-zero environment temperature $T$. This can be seen by computing the purity of the state, which, as summarized in Appendix~\ref{appendix: Gaussian states}, is equal to $P({\bm\sigma}^{\rm (out)}_{(n)})=1/\sqrt{{\rm det}{\bm\sigma}^{\rm (out)}_{(n)}}$. A quick calculation produces $P({\bm\sigma}^{\rm (out)}_{(n)})=(1+2n_{\text{B}})^{-2}$. The
mixed nature of the out state (\ref{outstate}) implies, in particular, that entanglement entropy 
(which quantifies entanglement only when the state of the total system is pure)
is not an appropriate  measure to quantify the entanglement between modes $(n, -n)$. 

Instead, we use logarithmic negativity ($E_{\mathcal{N}}$), defined in Appendix~\ref{appendix: Gaussian states}, which is based on the Peres-Horodecki or PPT 
(Positivity of the Partial Transpose) 
criterion~\cite{Peres:1996dw,Plenio:2005cwa,Simon:1999lfr}. For Gaussian states, $E_{\mathcal{N}}$  can be computed from the symplectic eigenvalues of the ``partially transposed'' covariance matrix.
 Furthermore, for the systems we are interested in here---single mode subsystems and Gaussian states---$E_{\mathcal{N}}$ is a faithful entanglement quantifier, in the sense that $E_{\mathcal{N}}$ is different from zero {\em if and only if} the state is entangled. It is also an entanglement monotone---higher $E_{\mathcal{N}}$ means more entanglement. $E_{\mathcal{N}}$ is measured in ebits or entangled bits, defined as the amount of entanglement contained in a Bell pair.

Applying expression \eqref{logneg} 
to the state (\ref{outstate}), we obtain
\bea\label{LogNeg}
&&\hspace{-.25cm}
E_{\mathcal{N}}[n] = \text{Max}\Big(0,-\log_{2}\Big[(1+2n_{\text{B}})(|\alpha_{n}|-|\beta_{n}|)^{2}\Big]\Big) \\
&&\hspace{-.25cm}= \text{Max}\Big(0,-\log_{2}\Big[(1+2n_{\text{B}})\left(\sqrt{1+|\beta_{n}|^2}-|\beta_{n}|)^{2}\right)\Big], 
\nonumber 
\eea
where we have used the identity $|\alpha_{n}|^{2}-|\beta_{n}|^{2}=1$ in the last equality. 
Note that the argument of the logarithm here is the minimum symplectic eigenvalue of 
the partially-transposed covariance matrix.

\begin{figure}[h!]
	\begin{center}
	\includegraphics[width=1.0\columnwidth]{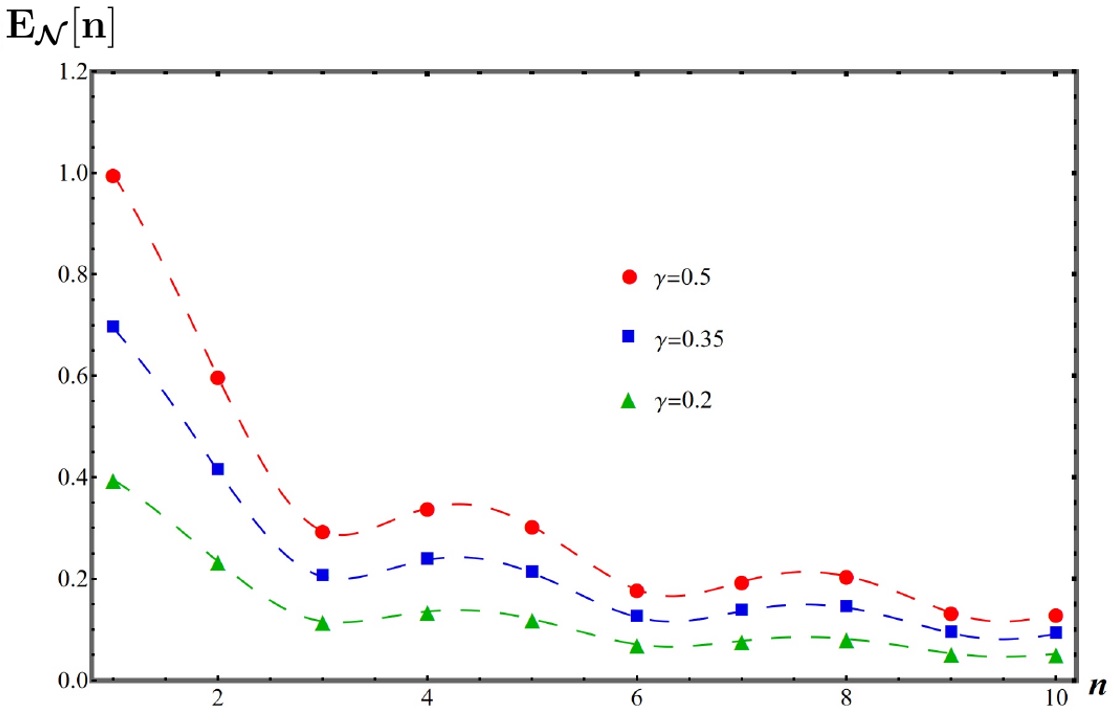}
	\end{center}
	\caption{(Color Online) Logarithmic Negativity between phonons labeled by $n$ and $-n$,  for zero ambient temperature $T=0$,  for three values of the quantum pressure
 parameter: $\gamma=0.2$ ({\em {green triangles}\/}), $\gamma=0.35$ ({\em {blue squares}\/}) and 
 $\gamma=0.5$ ({\em {red circles}\/}). The  mode index $n$ is discrete; we have added a continuous dashed line to increase the visibility of  the overall dependence of $E_{\mathcal{N}}$ with $n$. For this plot, all other parameters are the same as in Fig.~\ref{fig1}.}
	\label{Entanglement in Mode Space}
\end{figure}

We first discuss the situation in which the ambient temperature is zero. Substituting  $n_B=0$
in Eq.~(\ref{LogNeg}), we obtain $E_{\mathcal{N}}[n] =\text{Max}\Big(0,-\log_{2}\Big[\left(\sqrt{1+|\beta_{n}|^2}-|\beta_{n}|)^{2}\right)\Big]$. This expression is equal to zero if and only if $\beta_n=0$, and grows monotonically with $\beta_n$. This result tells us, on the one hand, that the members  within each created phonon-pair are entangled. Furthermore, since, for $T=0$, $|\beta_n|^2$ is equal to the number of pairs created, the total entanglement grows monotonically with the number of pairs created.  
Fig.~\ref{Entanglement in Mode Space} shows $E_{\mathcal{N}}$ versus the mode index $n$, for the expansion history given in \eqref{Eq:expansionchoice},  where we can recognize the  shape of $|\beta_n|^2$ shown in Fig.~\ref{fig1}, namely a rapid fall-off with increasing $n$ and oscillations dictated by the angular size of the horizon.

\begin{figure}[ht]
        \begin{center}
	\includegraphics[width=0.48\textwidth]{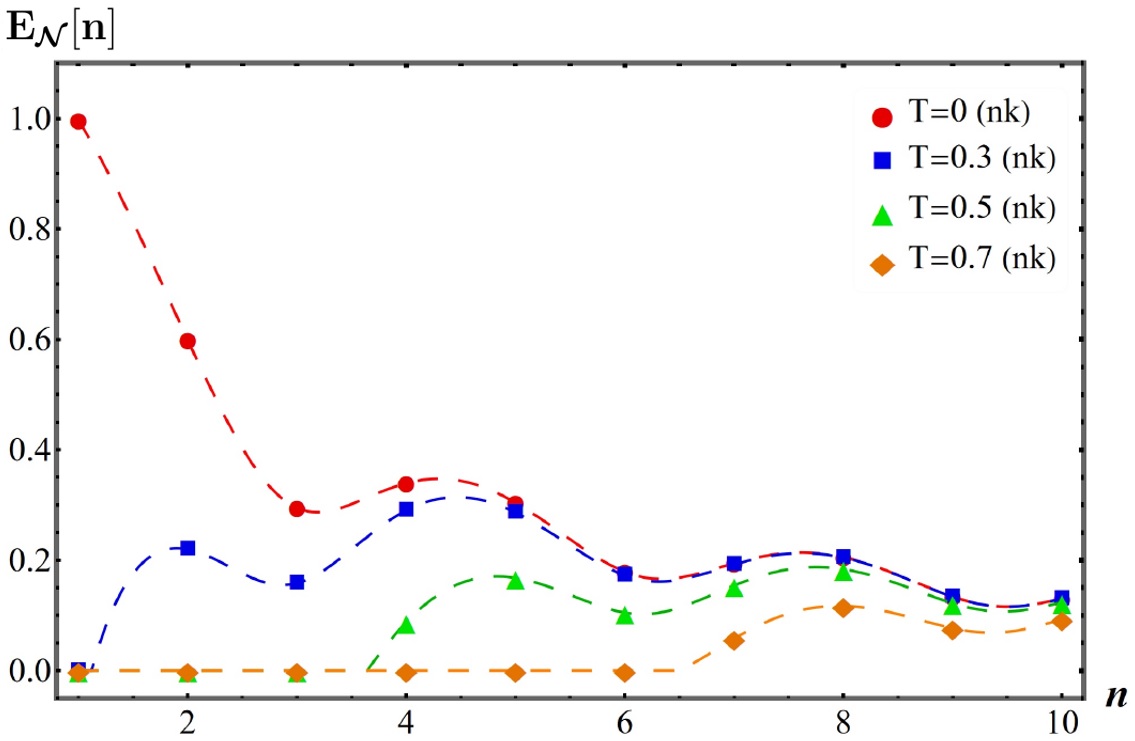}
        \end{center}
	\caption{(Color Online) Logarithmic negativity $E_{\mathcal{N}}[n]$ versus the mode index $n$, for various values of the environment temperature,  $T=0$ nK ({\em {red circles}\/}), 
 $T=0.3$ nK ({\em {blue squares}\/}), $T=0.5$ nK ({\em {green triangles}\/}), and $T=0.7$ nK ({\em {orange diamonds }\/}). The figure shows that  entanglement in the final state is degraded by ambient thermal noise, and that the entanglement  in pairs with small $n$ is more fragile. For this plot, we use quantum pressure  $\gamma=0.5$,
 with all other parameters being the same as in Fig.~\ref{fig1}.}
	\label{Entanglement Degradation Thermal Noise}
\end{figure}

\begin{figure}[ht]
	\centering
	\includegraphics[width=0.48\textwidth]{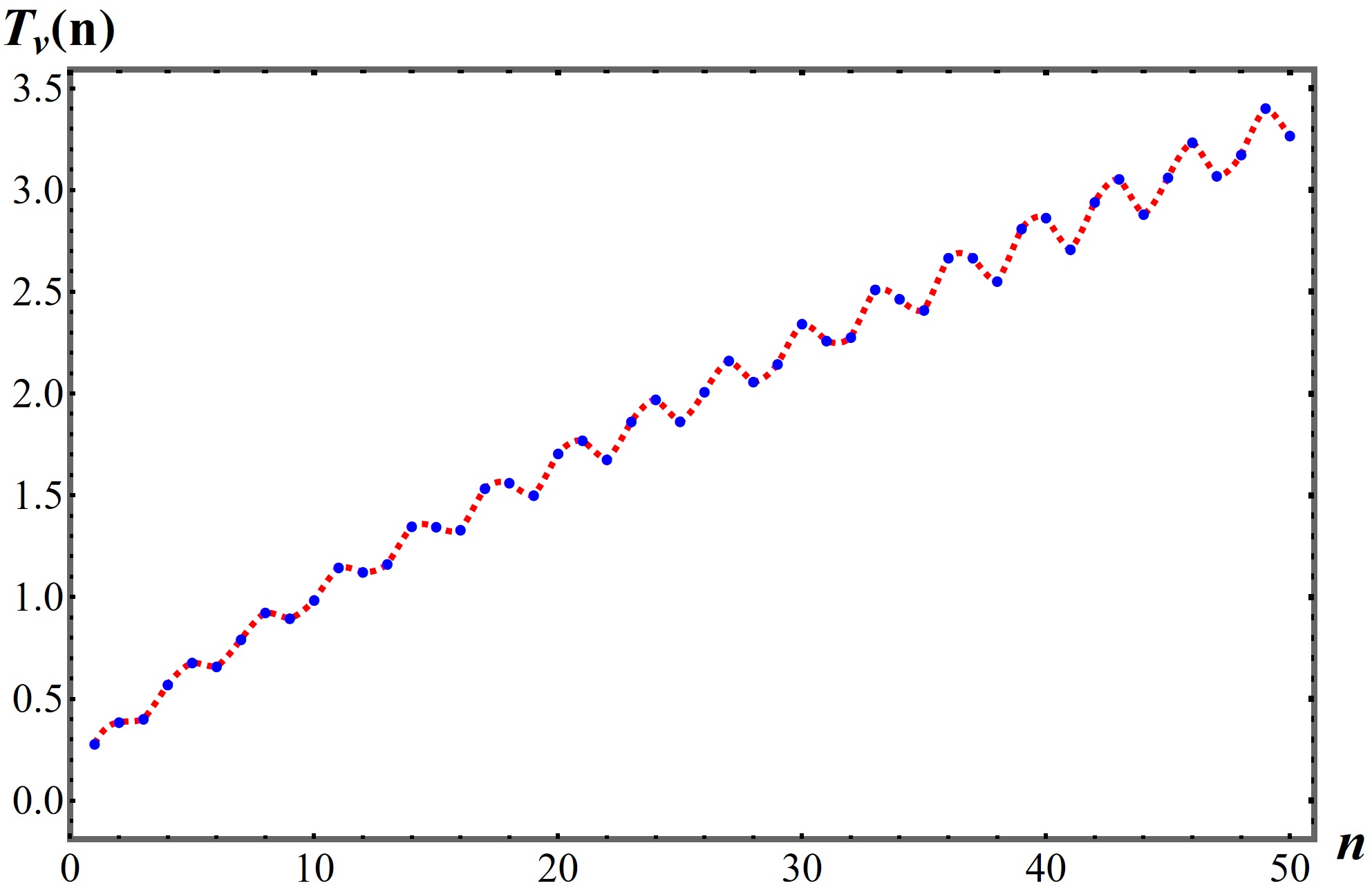}
	\caption{(Color Online) Critical temperature  $T_{v}(n)$ (in nK) at which the entanglement in the mode pair $(n,-n)$ completely vanishes. The blue dots represent the data for each mode index and the continuous red dashed line shows the overall dependence of $T_{v}$ with $n$. Larger ambient temperature is needed to degrade the entanglement  in pairs with large mode index $n$. The oscillations originate in the oscillatory character of  $|\beta_n|^2$. For this figure, we consider the same parameters as in Fig.~\ref{Entanglement Degradation Thermal Noise}.} 
	\label{TcrFIG}
\end{figure}

The second important lesson from expression \eqref{LogNeg} is the effect of ambient thermal noise on the generation of entanglement. This expression tells us that $E_{\mathcal{N}}$ is zero when the argument of the logarithm is larger than one. The effect of thermal noise is to add the factor $(1+2n_B)$ (recall, $n_B$ is the mean number of thermal quanta). This factor is larger than one, and we can always make $E_{\mathcal{N}}$ equal zero by increasing $n_B$. In other words, {\em thermal noise degrades the entanglement between modes $n$ and $-n$}, even making it disappear completely above a threshold temperature, which we will denote by $T_{v}(n)$. The value of this critical temperature can be easily obtained from expression \eqref{LogNeg}, by writing $n_B$ in terms of $T_{v}(n)$.
We obtain\footnote{The same result for $T_{v}(n)$ can be  obtained from the Bell-like inequality $\Delta<0$,
with $\Delta$ defined in Eq.~(\ref{eq:deltaDef}), as explained in Appendix~\ref{appendix: Gaussian states}.}
\begin{equation}\label{Tcr}
T_{v}(n)=\frac{E_n}{k_B} \left[ {\rm ln} \Biggl\{ 1+\frac{1}{|\beta_n|(|\beta_n|+\sqrt{1+|\beta_n|})} \Biggr\} \right]^{-1},
\end{equation}
The interpretation of this expression is simple: when thermal noisy phonon quanta dominate over the quanta created by the expansion, the entanglement in  the pair $(n,-n)$ vanishes. In other words, the entanglement in the final state results from a competition between the pair-creation and the environmental noise. But recall that the number of created pairs is dictated by $|\beta_n|^2$, which falls off approximately as $n^{-2}$, while the number of thermal quanta $n_B$ falls exponentially fast with $n$. Hence, thermal noise will degrade more easily the entanglement in pairs with lower value of the mode index $n$. Indeed, this is shown in Fig.~\ref{Entanglement Degradation Thermal Noise}, where we plot $E_{\mathcal{N}}$ versus $n$ for different ambient temperatures, showing that $E_{\mathcal{N}}[n]$ is
more strongly suppressed at small mode index $n$.  
This behavior is also exhibited in  Fig.~\ref{TcrFIG}, which shows that the threshold temperature $T_{v}$ 
at which $E_{\mathcal{N}}[n]$ approximately increases
(with small oscillations) with increasing $n$.  We 
emphasize again that, while the method is general, the results plotted in Figs.~\ref{Entanglement in Mode Space},
\ref{Entanglement Degradation Thermal Noise},
\ref{TcrFIG}
apply specifically to the expansion history \eqref{Eq:expansionchoice}. 

In summary, in the idealized situation of no initial density fluctuations, the spontaneous phonon-pair creation comes together with the generation of entanglement. We find that the presence of thermal noise drastically changes this picture. The stimulated pair-creation dominates over the spontaneous emission. 
On the other hand, thermal noise degrades the entanglement in the final state, rendering the state classical. The expressions derived in this section quantify these quantum effects and their degradation.

\subsection{Losses and efficiency}\label{Section 3d}

The discussion until now assumes an ideal situation where the process of expansion is a quantum channel with no losses.
Thus, there is no decoherence, and the probes used to detect perturbations  in the final static BEC have perfect efficiency. Losses and inefficiencies are, however, ubiquitous in realistic situations, and the goal of this section is quantify the effects they have on the quantum coherence of the final state using a simple model.

\begin{figure}[h!]
	\begin{center}
	\includegraphics[width=1.0\columnwidth]{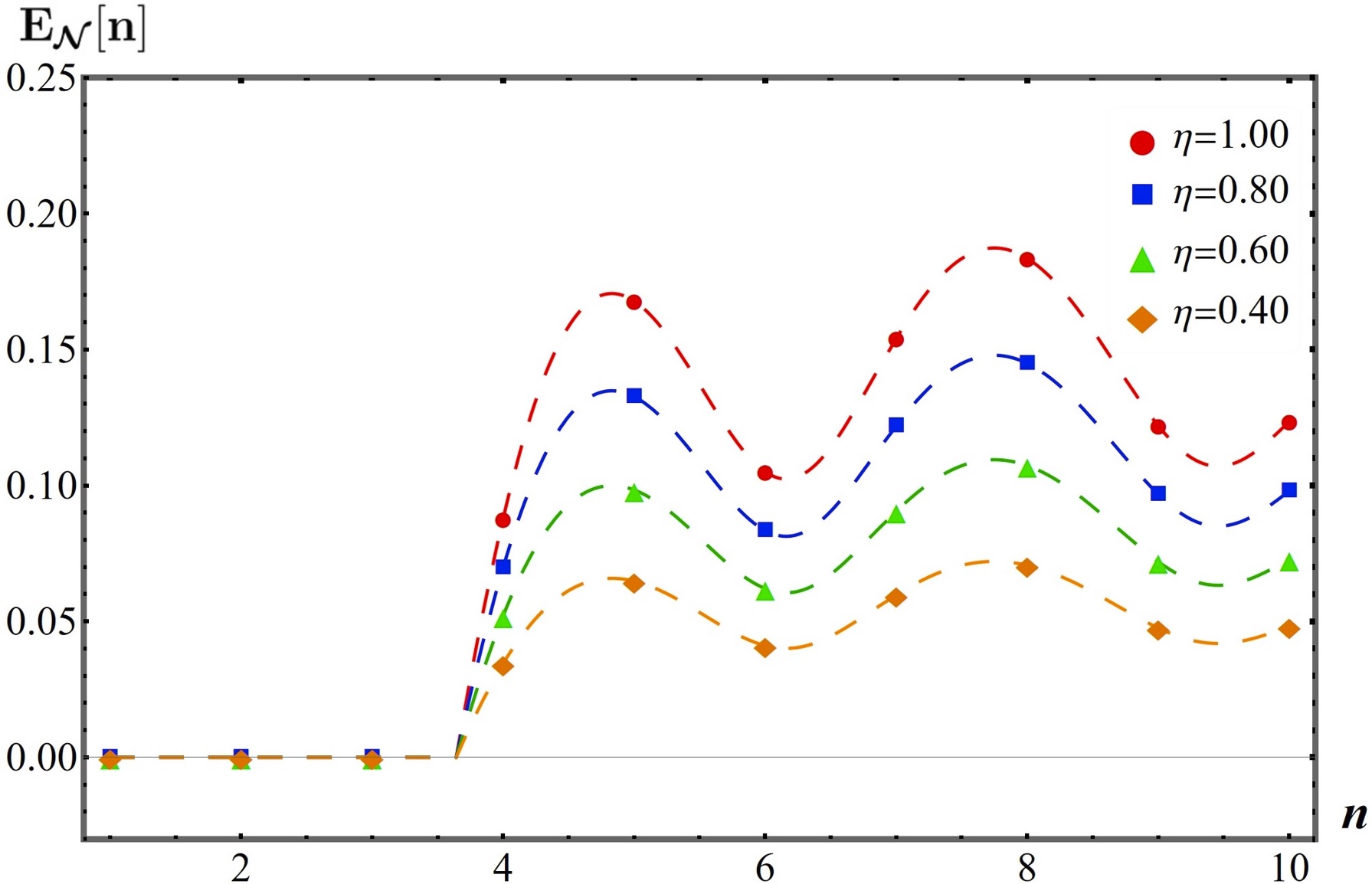}
	\end{center}
	\caption{(Color Online) Logarithmic negativity between phonons labeled by $n$ and $-n$, for various values of the efficiency parameter  
 $\eta=1$ ({\em {red circles}\/}), 
 $\eta=0.80$ ({\em {blue squares}\/}), $\eta=0.6$ ({\em {green triangles}\/}), and $\eta=0.4$ ({\em {orange diamonds}\/}). For this plot, we took the parameters in Fig.~\ref{fig1}, the environment temperature $T=\qty{0.5}{nK}$ and the quantum pressure to be $\gamma=0.5$.}
	\label{Thermal state & Losses}
\end{figure}

A simple yet useful model is provided by the so-called quantum attenuator channel  (see, e.g., Ref.~\cite{Serafini:2017}) characterized by a loss factor $\eta\in (0,1)$. This is a Gaussian channel, in the sense that the Gaussianity of the state is preserved. More concretely, the channel transforms the final state as
\be \Big(\bm{\mu}^{\rm (out)}_{(n)},\bm{\sigma}^{\rm (out)}_{(n)}\Big) \rightarrow \Big(\sqrt{\eta}\, \bm{\mu}^{\rm (out)}_{(n)},\, \eta\, \boldsymbol{\sigma}^{\text{out}}_{(n)}+(1-\eta)\, \mathbb{I}_{4}\Big)\, . \ee
Simply, quanta are detected with probability $\eta$ or are otherwise lost ($\eta=1$ corresponds to the ideal situation discussed above).
Following previous sections, we can re-evaluate the logarithmic negativity  $E_{\mathcal{N}}[n]$ for a pair of modes $(n,-n)$ and quantify entanglement in the final state after losses and/or inefficiencies.  We obtain 
\be \label{LNeta} E_{\mathcal{N}}[n]=\text{Max}\big(0,-\log_{2}\nu_{\text{T}, \eta}[n]\big)\, \ee
where
\begin{eqnarray}\label{Losses}
&& \hspace{-0.0cm}\nu_{\text{T},\eta}[n] = \frac{1}{4\sqrt{2}}\sqrt{X-\sqrt{Y}}, \ \ {\rm with} \nonumber \\
X & = & 16\bigg[2(1-\eta)^{2} + 2(1+2n_{\text{B}})^{2}\eta^{2}(|\alpha_{n}|^{4}+|\beta_{n}|^{4}) \nonumber \\
& + & 4(1+2n_{\text{B}})\eta(1-\eta)(|\alpha_{n}|^{2}+|\beta_{n}|^{2}) \nonumber \\
& + & 12\, (1+2n_{\text{B}})^{2}\eta^{2}|\alpha_{n}|^{2}|\beta_{n}|^{2}\bigg], \\
Y & = & (16\eta(1+2n_{\text{B}}))^{2}\bigg[ 64(1-\eta)^{2}|\alpha_{n}|^{2}|\beta_{n}|^{2} \nonumber \\
 &+&  64(1+2n_{\text{B}})^{2}\eta^{2}(|\alpha_{n}|^{4}+ |\beta_{n}|^{4})|\alpha_{n}|^{2}|\beta_{n}|^{2}  \nonumber \\
& + & 128(1+2n_{\text{B}})\eta(1-\eta)(|\alpha_{n}|^{2}+|\beta_{n}|^{2})|\alpha_{n}|^{2}|\beta_{n}|^{2} \nonumber \\
& + &  128(1+2n_{\text{B}})^{2}\eta^{2}|\alpha_{n}|^{4}|\beta_{n}|^{4} \bigg].
\end{eqnarray}
As a check, this expression reduces to Eq.~\eqref{LogNeg} obtained in the previous section in the limit $\eta\to 1$ (no losses). Furthermore,  $E_{\mathcal{N}}[n]$ vanishes when $\eta\to 0$, as expected, since in this limit none of the pairs created by the expansion get registered in the detectors. In between these limits, $E_{\mathcal{N}}[n]$ decreases monotonically when $\eta \to 0$. Unsurprisingly, losses and inefficiencies are sources of dissipation of quantum coherence or entanglement. Fig.~\ref{Thermal state & Losses} shows $E_{\mathcal{N}}[n]$ for different values of $\eta$, showing that more efficient detectors (i.e., with smaller
$\eta$) have more depletion of entanglement.

\subsection{Single mode squeezed states and inputs}\label{Section 3c}

In view of the fragility of quantum entanglement to thermal noise and losses, it is of interest to envisage mechanisms to amplify these quantum correlations. Such a mechanism would assist in keeping alive the quantum signature of the pair-creation process in situations where the conditions of the experiment would otherwise render the final state completely classical. 

We use a mechanism proposed in \cite{Agullo:2021vwj,Brady:2022ffk}, consisting of replacing the initial vacuum, before the expansion of the BEC ring, by a {\em single-mode squeezed state}. Such a state is separable, in the sense that it does not contain entanglement between any mode with index $n$. However, it contains initial quanta which 
stimulate the production of additional entangled pairs. This stimulated creation strengthens the entanglement in the final state, making it more robust under deleterious effects. However, this mechanism does not
create entanglement by itself: the final system entanglement would be zero if the BEC ring does not expand (since the initial excitations are unentangled), and in this sense the entanglement in the final state can all be attributed to the expansion. The initial squeezing acts as a ``catalyzer'' for the generation of entanglement.

The limitation of this strategy to amplify or catalyze the production of entanglement is the difficulty of generating phonons in single-mode squeezed states in the lab.  We postpone for future work the question of how a toroidal BEC experiment may realize a single-mode squeezed 
initial state, taking the point of view that 
the strategy presented in this section can be of interest as a concrete example for increasing the visibility of the quantum signatures of the pair-creation process.

We proceed to consider an initial phonon state for the mode-pair $(n,-n)$ corresponding to a single-mode squeezed state with thermal noise in it. This is a Gaussian state of the form (suppressing the mode index label)
\bea\label{Squeezed Covariance Matrix}
{\bm \mu}_r^{\rm (in)}&=&(0,0,0,0)^{\top}\, , \\
\boldsymbol{\sigma}^{\text{in}}_{r} &=& (1+2n_{\text{B}})\begin{bmatrix}
		e^{2r} & 0 & 0 & 0 \\
		0 & e^{-2r} & 0 & 0 \\
		0 & 0 & 1 & 0 \\
		0 & 0 & 0 & 1 \\
	\end{bmatrix}.
\eea
In this  state, the mode $-n$ is in a thermal state, while the mode $n$ is in a thermal-squeezed state with squeezing intensity $r\in \mathbb{R}$. The modes are uncorrelated. In the limit ${r}\to 0$ this state reduces to the thermal state used in the previous section. We have chosen the squeezing in a particular `direction'---namely, we have squeezed the state in the $\hat p_n$ direction, and anti-squeezed in $\hat x_n$ direction. The direction of the squeezing can be controlled by introducing a squeezing angle $\phi$. The results of this section are independent of the choice of $\phi$, so we will use $\phi=0$ for simplicity in the presentation. 

\begin{figure}[ht]
	\centering
	\includegraphics[width=0.48\textwidth]{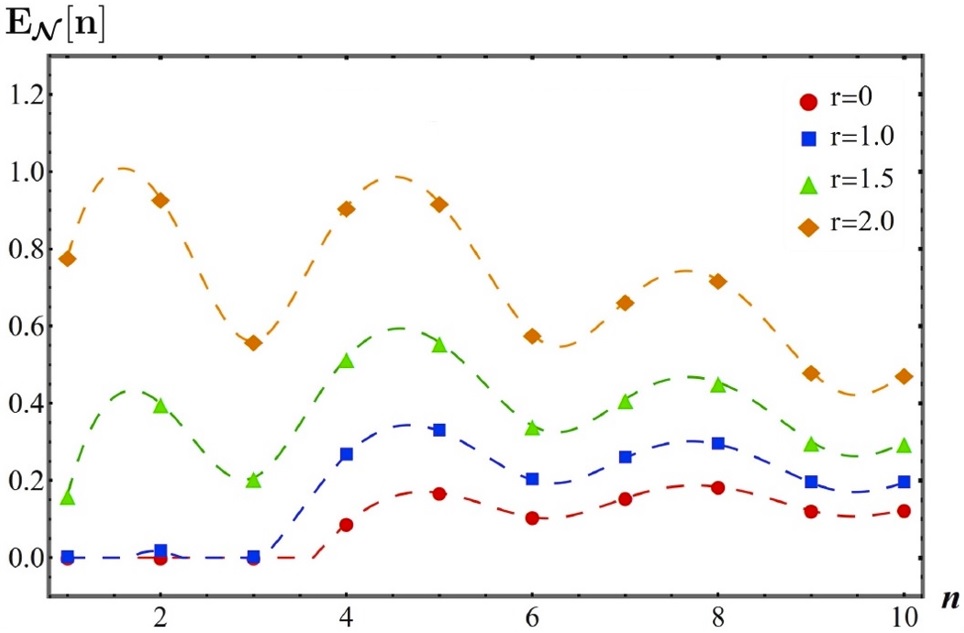}
	\caption{(Color Online) Logarithmic negativity $E_{\mathcal{N}}$$[n]$ for pairs $(n,-n)$,  versus the squeezing intensity $r$ and  ambient temperature $T=\qty{0.5}{nK}$. Various values of  initial squeezing $r$ are shown. This plot corresponds to quantum pressure  $\gamma=0.5$ and $\eta =1$, with the rest of the parameters
 being the same as in previous plots.}
	\label{Entanglement Enhancement Squeezing}
\end{figure}

Following the strategy used in the previous section, we can compute the covariance matrix of the final state as ${\bm \sigma}_r^{\rm (out)}={\bf S}\cdot{\bm \sigma}_r^{\rm (in)}\cdot{\bf S}^{\top}$, and compute from it the logarithmic negativity.  Note that we also include
losses parameterized by $\eta$ in this calculation. We obtain
\be E_{\mathcal{N}}[n]=\text{Max}\big(0,-\log_{2}\nu_{\text{T}, {r}, \eta}[n]\big)\, \ee
where
\begin{eqnarray}\label{Alternate Losses}
&&\nu_{\text{T},{r},\eta}[n] = \frac{1}{4\sqrt{2}}\sqrt{X_{r}-\sqrt{Y_{r}}},  \\
&&X_{r} = 16\bigg[2(1-\eta)^{2} + 2(1+2n_{\text{B}})^{2}\eta^{2}(|\alpha_{n}|^{4}+|\beta_{n}|^{4}) \nonumber \\
&& + 4(1+2n_{\text{B}})\eta(1-\eta)\cosh^{2}r(|\alpha_{n}|^{2}+|\beta_{n}|^{2}) \nonumber \\
&& + 4(1+2n_{\text{B}})^{2}\eta^{2}(1+2\cosh(2r))|\alpha_{n}|^{2}|\beta_{n}|^{2}\bigg],  \\
&&Y_{r} =  (16\eta(1+2n_{\text{B}}))^{2}\bigg[16\sinh^{4}r (1-\eta)^{2}(|\alpha_{n}|^{4}+|\beta_{n}|^{4}) \nonumber \\
&& + 4(1-\eta)^{2}\big(3+12\cosh(2r)+\cosh(4r)\big)|\alpha_{n}|^{2}|\beta_{n}|^{2} \nonumber \\
&& + 64(1+2n_{\text{B}})^{2}\eta^{2}\cosh^{2}r(|\alpha_{n}|^{4}+|\beta_{n}|^{4})|\alpha_{n}|^{2}|\beta_{n}|^{2}  \nonumber \\
&& + 128(1+2n_{\text{B}})\eta(1-\eta)\cosh^{4}r(|\alpha_{n}|^{2}+|\beta_{n}|^{2})|\alpha_{n}|^{2}|\beta_{n}|^{2}\nonumber \\
&& + 128(1+2n_{\text{B}})^{2}\eta^{2}\cosh^{2}r\cosh(2r)|\alpha_{n}|^{4}|\beta_{n}|^{4}\bigg]. 
\end{eqnarray}
These expressions reduce to Eqs.~\eqref{LNeta} and \eqref{Losses} obtained in the last section in the limit $r\to 0$. The physical content in these expressions is shown in Fig.~\ref{Entanglement Enhancement Squeezing}, which corresponds to a situation with no losses ($\eta=1$). This plot shows the way initial squeezing  amplifies the entanglement in the final state, and compensates for the deleterious effects of thermal noise. Fig.~\ref{Thermal state, Losses & Squeezing} shows how the entanglement degrades due to imperfect detectors even if we start with a thermal single-mode squeezed state of phonons.

To summarize, in this section, we have discussed that in order to quantify entanglement, we need to construct the covariance matrix which is basically a collection of all possible correlations between phonon modes. The PPT criterion and logarithmic negativity then help us quantify the amount of entanglement in the phonon pairs for different choices of the initial quantum state, i.e., vacuum, thermal, and single-mode squeezed states. In the next section, we propose a protocol that can measure these mode correlations, thereby experimentally revealing the entanglement structure of phonons in an expanding toroidal BEC. 

\begin{figure}[h!]
	\begin{center}
	\includegraphics[width=1.0\columnwidth]{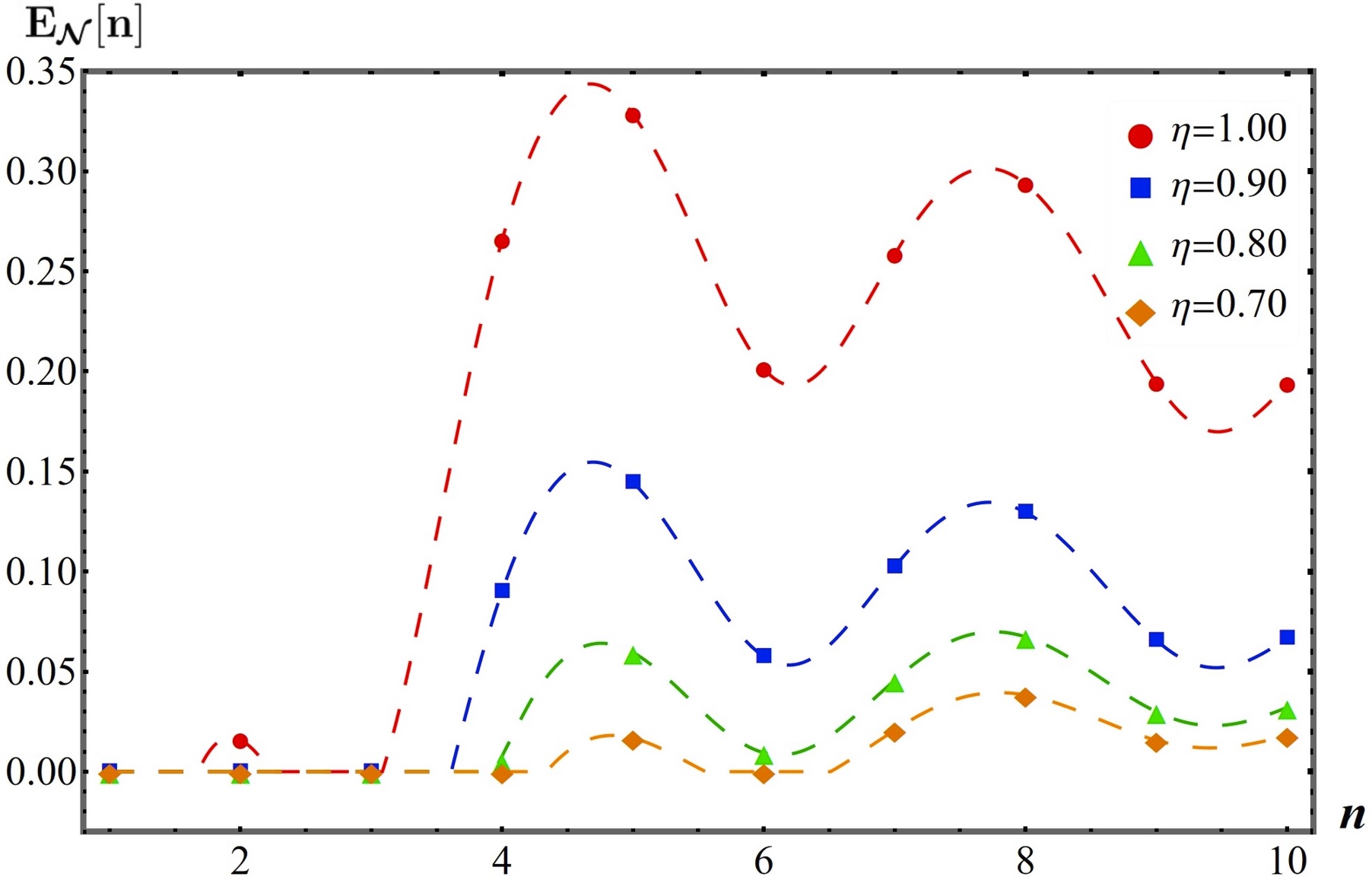}
	\end{center}
	\caption{(Color Online) Logarithmic negativity between phonons labeled by $n$ and $-n$, for various values of the efficiency parameter $\eta=1$ ({\em {red}\/}), $\eta=0.80$ ({\em {blue}\/}), $\eta=0.6$ ({\em {green}\/}) and $\eta=0.8$ ({\em {red}\/}. For this plot, we took the parameters in Fig.~\ref{fig1}, the environment temperature $T=\qty{0.5}{nK}$, the quantum pressure $\gamma=0.5$, and the single-mode squeezing parameter to be $r=1.0$.}
	\label{Thermal state, Losses & Squeezing}
\end{figure}

\begin{figure*}[ht]
	\centering
	\includegraphics[width=\textwidth]{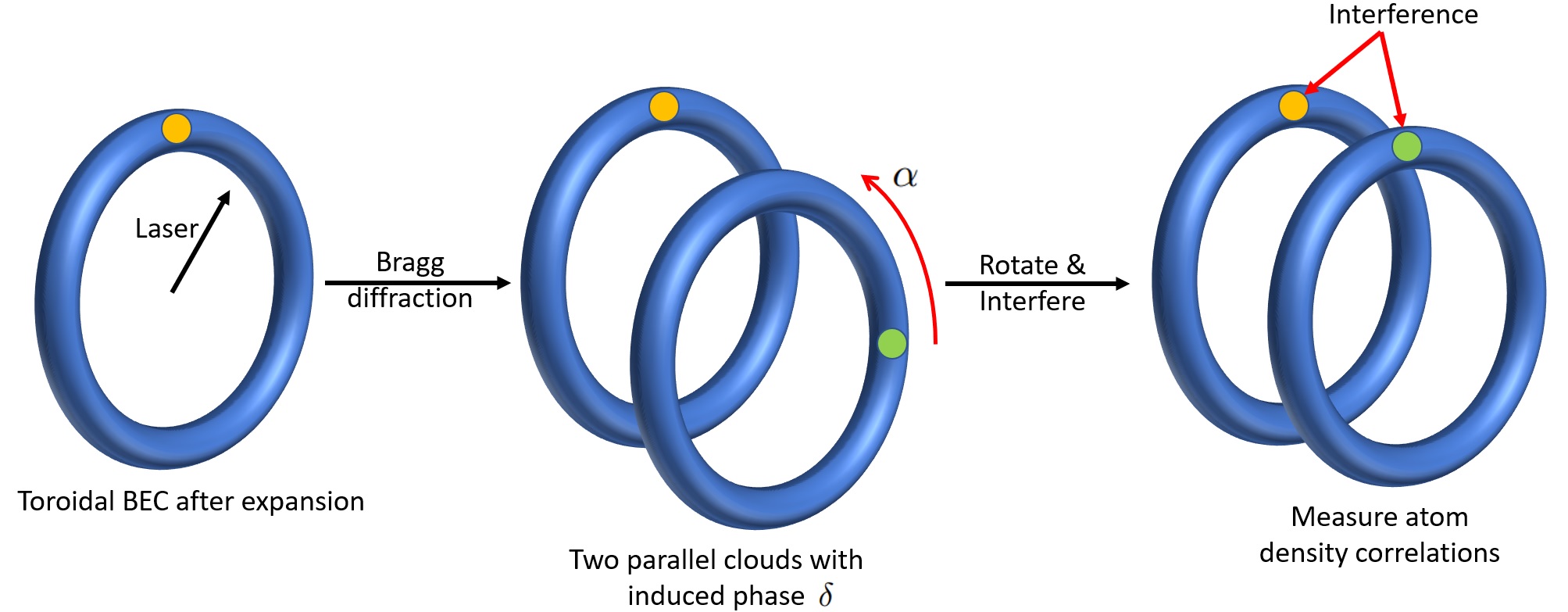}
	\caption{(Color Online) A schematic figure showing the protocol to measure mode entanglement in the BEC. The first stage (first panel) is where the expanded BEC is Bragg diffracted using a laser, which then splits into two parallel clouds with an induced phase difference of $\delta$. Then, in the second stage (second panel), one of the rings is rotated by some angle $\alpha$ which brings the two points (shown in green and orange colors) in front of each other. In the third stage (third panel), the two clouds are made to interfere with each other and the atom density correlations are obtained. This setup would allow
 the extraction of all types of correlations between phase and density fluctuations, and thus the covariance matrix can be built.}
	\label{Angular Hellweg}
\end{figure*}

\section{A protocol to measure entanglement}\label{Section 4}

Observing entanglement in realistic situations is a challenging task. In this section,
we discuss a possible strategy to achieve this goal adapted to the system we study in this article. This proposal contains several idealized ingredients, which could give rise to challenges in realistic situations. It nevertheless provides a concrete set of ideas which can be useful as the starting point of a more refined protocol adapted to concrete experimental setups. 

As discussed in Appendix~\ref{appendix: Gaussian states}, there is a trade off in the use of logarithmic negativity compared to simpler entanglement witnesses, such as the Bell-like inequality $\Delta <0$
(with $\Delta$ defined in Eq~(\ref{eq:deltaDef})). While $\Delta$ involves only a few moments of the final state, it only indicates whether entanglement is present in the final state in certain circumstances. More precisely, the limitations of this inequality are two: (i) it is not faithful, in the sense that 
$\Delta \geq 0$ does not rule out the existence of entanglement, and (ii) 
it
is not a quantifier, because a stronger violation of the inequality does not imply more entanglement. On the other hand, logarithmic negativity is a faithful quantifier for the simple systems we are interested in, namely Gaussian states and single-mode subsystems. The trade off is that its evaluation requires  knowledge of the entire covariance matrix of the final state, which amounts to having full knowledge of the state.  

To measure the elements of the covariance matrix of an inflationary toroidal BEC, we propose a generalization of the method due to Hellweg et al.~\cite{Hellweg:2003}, who measured the phase correlation function of
a trapped BEC via a scheme that is analogous to the well-known stellar
interferometry measurements of Hanbury-Brown and Twiss~\cite{Hanbury-Brown & Twiss 1,Hanbury-Brown & Twiss 2,Hanbury-Brown & Twiss 3}.
Hellweg et al.\ accomplished this in two steps: First, by
using Bragg diffraction to split the BEC into two identical copies (with a controllable
phase difference between them);  importantly, these two copies were shifted spatially
relative to each other.  The second step is to interfere the two separated condensates, measuring density correlations in the final system.  The result of this final
measurement  reflects {\em phase\/} correlations of the original BEC.  

In generalizing this scheme to the present case, it is not sufficient to 
merely extract phase correlations.  Indeed, we also require density-density and
phase-density correlations, as a function of angle, i.e., we need the correlation functions
\begin{eqnarray}
C_{\hat n_1\hat n_1}(\alpha) & := & \langle\{\hat{n}_{1}(\alpha),\hat{n}_{1}(0)\}\rangle, \\
C_{\hat \phi_1\hat \phi_1}(\alpha) & := & \langle\{\hat{\phi}_{1}(\alpha),\hat{\phi}_{1}(0)\}\rangle, \\
C_{\hat n_1\hat \phi_1}(\alpha) & := & \langle\{\hat{n}_{1}(\alpha),\hat{\phi}_{1}(0)\}\rangle,
\end{eqnarray}
where $\alpha$ is the separation angle.   From these functions, we can obtain the correlations between modes $(n,-n)$ by a simple Fourier transform, as we discuss below.  Since a Gaussian state is completely 
determined by the set of quadratic correlation functions, it makes sense that these three functions
are sufficient to reproduce the full covariance matrix.  Mathematically, the connection comes
from Eqs.~(\ref{eq:phianddensity}) that relate density and phase fluctuations to the 
mode operators ${\hat a}_n^\dagger$ and ${\hat a}_n$, and via Eq.~(\ref{eq:connection}) that connects the
mode operators to the $\hat{x}_n$, $\hat{p}_n$.  

To see how these correlation functions could be measured, we now describe a natural
generalization of the Hellweg et al.\ steps to the present case of an 
 expanding toroidal BEC, as illustrated in Fig.~\ref{Angular Hellweg}.
The first
step assumes it is possible to use Bragg diffraction to split the original toroidal BEC into two
identical copies, with an induced phase $\delta$ between them (that is 
experimentally controllable).  We also
assume the resulting clouds can be rotated an angle $\alpha$ relative to each other, 
after which they are allowed to 
 evolve freely for some time.  Finally, the clouds are interfered, with density correlations
 being measured in the final system.  The preceding steps can be encapsulated in the following
 expression for the final condensate field operator as a function of angle $\theta$:
\begin{eqnarray}
\hat{\Phi}(\theta) & = & \frac{1}{2}\Big[ \sqrt{n_{0}+\hat{n}_{1}(\theta-\alpha/2)}e^{i\big(\phi_{0}+\hat{\phi}_{1}(\theta-\alpha/2)\big)} \nonumber \\
& + & e^{i\delta}\sqrt{n_{0}+\hat{n}_{1}(\theta+\alpha/2)}e^{i\big(\phi_{0}+\hat{\phi}_{1}(\theta+\alpha/2)\big)} \Big],
\label{Eq:fieldoperator}
\end{eqnarray}
which also depends on the controllable angular displacement $\alpha$ and controllable
phase difference $\delta$.  Here, we have made use of the Madelung representation $\hat{\Phi}(\theta)=\sqrt{\hat{n}}\, e^{i\hat{\phi}}$, with
$\hat{n}=n_{0}+\hat{n}_{1}(\theta)$ and $\hat{\phi}=\phi_{0}+\hat{\phi}_{1}(\theta)$, as described in Eqs.~(\ref{Madelung representation}) and (\ref{perturbations}).  Eq.~(\ref{Eq:fieldoperator})
is precisely an angular version of Eq.~(5) of Hellweg et al.~\cite{Hellweg:2003}.

Now we show that a measurement of the condensate density two-point correlation function at 
coincident points,  $\langle\hat{N}(\theta)\hat{N}(\theta)\rangle_{\alpha,\delta}$,   where $\hat{N}(\theta)=\hat{\Phi}^{\dagger}(\theta)\hat{\Phi}(\theta)$, contain the correlations we are looking for, as we describe now---the labels $\alpha $ and $\delta$ remind us about the rotation angle $\alpha$ and induced Bragg phase $\delta$ chosen in the procedure.  Let us focus on an arbitrary point of the final cloud, which, without loss of generality,  we can choose as $\theta=0$.  Then, the symmetric correlator   $\langle\{\hat{N}(0),\hat{N}(0)\}\rangle_{\alpha,\delta}$  obtained is related to  phase $\hat{\phi}_{1}(\theta)$ and density $\hat{n}_{1}(\theta)$ two-point correlators by 
\begin{eqnarray}\label{CondensateDensityCorrelation}
&& \langle\{\hat{N}(0),\hat{N}(0)\}\rangle_{\alpha,\delta} = \frac{1}{2}n^{2}_{0}(1+\cos\delta)^{2} \nonumber \\
& + & \frac{1}{16}\Big[\tilde{C}_{\hat{n}_1\hat{n}_1}(\alpha)\ (1+\cos\delta)^{2} + 
\tilde{C}_{\hat{\phi}_1\hat{\phi}_1}(\alpha)\, 4n^{2}_{0}\, \sin^{2}\delta\Big].
\end{eqnarray}
where we have defined 
\begin{eqnarray}\label{DensityTildeCorrelator}
\tilde{C}_{\hat n_1\hat n_1}(\alpha) & \equiv & -2\Big(C_{\hat n_1\hat n_1}(\alpha)-C_{\hat n_1 \hat n_1}(0)\Big), \\ \label{PhaseTildeCorrelator}
\tilde{C}_{\hat \phi_1\hat \phi_1}(\alpha) & \equiv & -2\Big(C_{\hat \phi_1\hat \phi_1}(\alpha)-C_{\hat \phi_1\hat \phi_1}(0)\Big).
\end{eqnarray}
Therefore, by measuring  $\langle\{\hat{N}(0),\hat{N}(0)\}\rangle_{\alpha,\delta}$ for two different induced phases, say $\delta_{1}$ and $\delta_{2}$, and fixing the relative rotation angle $\alpha$, the above linear equation can be solved for $\tilde{C}_{\hat n_1\hat n_1}(\alpha)$ and $\tilde{C}_{\hat \phi_1\hat \phi_1}(\alpha) $. From these correlations it is simple to obtain the correlations functions we are actually interested in---$C_{\hat n_1\hat n_1}(\alpha)$ and $C_{\hat \phi_1\hat \phi_1}(\alpha)$---by noticing that 
\begin{equation} \label{A}
\frac{1}{4\pi}\int_{0}^{2\pi}d\alpha~C_{\hat n_1\hat n_1}(\alpha) = 0\, .
\end{equation}
This is because $C_{\hat n_1\hat n_1}(\alpha)$
does not include  the zero mode $n=0$ in its Fourier series, since this homogeneous mode has been absorbed in the background condensate $n_0$. Thus, integrating (\ref{DensityTildeCorrelator}) over all possible rotation angles $\alpha$, and making use of \eqref{A}, produces $2\,C_{\hat n_1 \hat n_1}(0)$. 
The same argument applies to $C_{\hat \phi_1\hat \phi_1}(\alpha)$. 
Thus, the elementary density-density and phase-phase correlations can be extracted as follows:
\begin{eqnarray}\label{DensityCorrelationExtracted}
\hspace{-0.5cm} C_{\hat n_1\hat n_1}(\alpha) & = & -\frac{1}{2}\tilde C_{\hat n_1\hat n_1}(\alpha) + \frac{1}{4\pi}\int_{0}^{2\pi}d\alpha~\tilde{C}_{\hat n_1\hat n_1}(\alpha), \\ \label{PhaseCorrelationExtracted}
\hspace{-0.5cm} C_{\hat \phi_1\hat \phi_1}(\alpha) & = & -\frac{1}{2}\tilde{C}_{\hat \phi_1\hat \phi_1}(\alpha) + \frac{1}{4\pi}\int_{0}^{2\pi}d\alpha~\tilde{C}_{\hat \phi_1\hat \phi_1}(\alpha).
\end{eqnarray}
[Note that $C_{\hat n_1\hat n_1}(\alpha)$ can be directly measured in the original toriodal BEC, without
the interferometry steps; so, as a check, one could compare such a direct observation with the result obtained from \eqref{DensityCorrelationExtracted}.]

It remains to determine the mixed density-phase correlations $C_{\hat n_1\hat \phi_1}(\alpha)$. Since $\hat \phi_1$ and $\hat n_1$ are related by  $\hat n_1=-\frac{\hbar\mathcal{V}}{U}\frac{d}{dt}\hat \phi_1$, the missing correlation function can be obtained by taking time derivative of $C_{\hat \phi_1\hat \phi_1}(\alpha)(t)$. This requires a repetition of the above steps for different times after expansion $t>t_{f}$, in discrete time intervals.  That is, after 
expansion, Bragg diffraction, rotation and interference, we require a measurement of the correlations at time $t= t_f$ and a measurement of the
same for an identically prepared condensate at $t=t_{f}+\Delta t$.  The time interval $\Delta t$ should be small compared to any other time scales in the problem, such as the frequency associated with the smallest mode $\omega_{n=1}^{\text{f}}\equiv c/R(t_{\text{f}})$, or the time scale associated with the damping of density fluctuations, as was seen in {\cite{Eckel:2017uqx}}. This procedure yields the time evolution data for the phase-phase correlations, from which we can obtain the mixed correlation function: 
\begin{equation}\label{Time Derivative & Mixed Correlations}
C_{\hat n_1\hat \phi_1}(\alpha) \equiv \langle\big\{\hat{n}_{1}(\alpha,t),\hat{\phi}_{1}(0,t)\big\}\rangle = -\frac{1}{2}\frac{\hbar\mathcal{V}_{f}}{U}\frac{d}{dt}C_{\hat \phi_1\hat \phi_1}(\alpha,t), \nonumber
\end{equation}
where we have defined $\mathcal{V}_{f}\equiv\mathcal{V}(t_{f})$, and have made use of Eq.~(\ref{Mode Expansion for Initial Density}) that relates the density operator to the time derivative of the phase operator.

Once the two-point correlations in real space are known,  we can obtain the covariance matrix of the final state as follows. First, from $C_{\hat n_1\hat n_1}(\alpha)$, $C_{\hat \phi_1\hat \phi_1}(\alpha) $ and $ C_{\hat n_1\hat \phi_1}(\alpha) $ we can obtain the symmeterized second moments of the creation and annihilation operators as follows:
\begin{eqnarray}\label{ModeSpaceCorrelators}
&& \langle\{\hat{a}^{\rm (out)}_{n},\hat{a}^{\rm (out)}_{-n}\}\rangle = p_{f} \int_{0}^{2\pi}d\alpha~e^{-in\alpha} \bigg[(\chi^{*}_{n})^{2}C_{\hat n_1\hat n_1}(\alpha) \nonumber \\
&& + \mathcal{V}^{2}_{f}(\eta^{*}_{n})^{2}C_{\hat \phi_1\hat \phi_1}(\alpha) - 2\mathcal{V}_{f}\chi^{*}_{n}\eta^{*}_{n}C_{\hat n_1\hat \phi_1}(\alpha)\bigg], \nonumber\\ 
&& \langle\{\hat{a}^{\rm (out)}_{n},\hat{a}^{{\rm (out)}\, \dagger}_{n}\}\rangle =  -p_{f} \int_{0}^{2\pi}d\alpha~e^{-in\alpha} \nonumber \\
&& \times \bigg[|\chi_{n}|^{2}C_{\hat n_1\hat n_1}(\alpha) + \mathcal{V}^{2}_{f}|\eta_{n}|^{2}C_{\hat \phi_1\hat \phi_1}(\alpha)\bigg],
\end{eqnarray}
where we have defined the proportionality constant to be $p_{f}=-a^{1+2\gamma}_{f}\frac{U}{\hbar\mathcal{V}_{f}}$, and we have suppressed the term `out' in the mode functions for phase $\chi^{\rm (out)}_{n}=(2\omega^{f}_{n}a_{f}^{1+\gamma})^{-1/2}e^{-i\omega^{f}_{n}(t-t_{f})}$, and density $\eta^{\rm (out)}_{n}=-i\frac{\hbar}{U}\sqrt{\frac{\omega^{f}_{n}}{2a_{f}^{1+\gamma}}}e^{-i\omega^{f}_{n}(t-t_{f})}$. The rest of the
independent second moments, $\langle\{\hat{a}^{\rm (out)}_{n},\hat{a}^{\rm (out)}_{n}\}\rangle$ and $\langle\{\hat{a}^{\rm (out)}_{n},\hat{a}^{{\rm (out)}\, \dagger}_{-n}\}\rangle$, are zero.

With these moments we have full information of the covariance matrix of the final state in the creation and annihilation variables
\be {\bm \sigma}^{\rm out}_{({\bf A}, n)}=\langle \{  \hat{{\bf A}}^{\rm (out)}_{n}, \hat{{\bf A}}^{\rm (out)}_{n}\}\rangle \, , \ee
out of which we obtain the covariance matrix we are looking for as ${\bm \sigma}^{\rm out}_{(n)}={\bf B}\cdot {\bm \sigma}^{\rm out}_{({\bf A}, n)}\cdot {\bf B}^{-1}$, where the change-of-basis matrix $\bf B$ is given in Eqn.~\eqref{B}. 
Then, following the procedure outlined in Sec.~\ref{Section 3}, we can quantify the quantum correlations in this system via the logarithmic negativity $E_{\mathcal{N}}[n]$, that determines the amount of entanglement in the phonon mode pairs $(n,-n)$ in the expanding toroidal BEC.

The preceding analysis also gives us a hint as to what types of quantum correlations in the toroidal BEC lead to more entanglement, i.e., larger logarithmic negativity. Following the discussions in Sec.~\ref{sec:entanglement} and Appendix~\ref{appendix: Gaussian states}, we note that 
in the present two-mode case
the covariance matrix is a $4\times 4$ matrix, so that there are two symplectic eigenvalues of
the partially-transposed covariance matrix.  Since only the minimum 
of these two (which we call $\nu_{\text{min}}$) can be less
than unity (and contribute to the logarithmic negativity), it is sufficient to focus on 
$\nu_{\text{min}}$, which can be written in terms of the mode space correlators given in Eq.~(\ref{ModeSpaceCorrelators}):
\begin{equation}
\nu_{\text{min}} = \langle\{\hat{a}^{\rm (out)}_{n},\hat{a}^{{\rm (out)}\, \dagger}_{n}\}\rangle - |\langle\{\hat{a}^{\rm (out)}_{n},\hat{a}^{\rm (out)}_{-n}\}\rangle|.
\end{equation}
Since the logarithmic negativity $E_{\mathcal{N}}[n]$ is related to the logarithm of $\nu_{\text{min}}$,  in order to maximize entanglement, we need this eigenvalue to approach zero. Using Eq.~\eqref{ModeSpaceCorrelators}, we can express this condition in terms of real space correlators:
\begin{equation}\label{Lambda-Min}
\nu_{\text{min}} = -A_{+} -\sqrt{A_{-}^{2}+B^{2}},
\end{equation}
where the functions $A_\pm$ and $B$ are defined as follows:
\begin{eqnarray}
A_{\pm} & = & |\chi_{n}|^{2}C_{\hat n_1\hat n_1}(n) \pm |\eta_{n}|^{2}C_{\hat \phi_1\hat \phi_1}(n), \\
B & = & 2|\chi_{n}||\eta_{n}| \, C_{\hat n_1\hat \phi_1}(n),
\end{eqnarray}
where 
we define the Fourier transforms of the real-space correlators,
\begin{eqnarray}
C_{\hat n_1\hat n_1}(n) & = & p_{f}\int_{0}^{2\pi}d\alpha~e^{-in\alpha}C_{\hat n_1\hat n_1}(\alpha), \\
C_{\hat \phi_1\hat \phi_1}(n) & = & p_{f}\mathcal{V}_{f}^{2}\int_{0}^{2\pi}d\alpha~e^{-in\alpha}C_{\hat \phi_1\hat \phi_1}(\alpha), \\
C_{\hat n_1\hat \phi_1}(n) & = & p_{f}\mathcal{V}_{f}\int_{0}^{2\pi}d\alpha~e^{-in\alpha}C_{\hat n_1\hat \phi_1}(\alpha).
\end{eqnarray}
To maximize entanglement, we need $\nu_{\text{min}}$ to be small. Setting Eq.~(\ref{Lambda-Min}) to vanish, putting the square root on one side, and squaring both sides leads to:
\begin{equation}
C_{\hat{n}_1\hat{n}_1}(n)C_{\hat{\phi}_1\hat{\phi}_1}(n) = \Big(C_{\hat n_1\hat \phi_1}(n)\Big)^{2},
\end{equation}
a condition on the correlation functions, as a function of mode index, that maximizes 
$E_{\mathcal{N}}[n]$.

\section{Concluding Remarks}\label{Section 5}
\label{sec:concl}

This paper provides a quantitative analysis of the generation of quantum entanglement in the process of pair-creation in a thin toroidal BEC with a time-dependent radius. This system constitutes  an analog simulator for the behavior of quantum fields propagating in an expanding universe, and in particular during inflation. Such expanding BEC rings have been experimentally produced  \cite{Eckel:2017uqx,Banik:2021xjn}, where the red-shift of density perturbations induced by the expansion and an analog of the process of re-heating have been observed. The phonon pair-production phenomenon in this system, triggered by the expansion, was then studied theoretically in Ref.\cite{Bhardwaj:2020ndh}, using a model based on the Bogoliubov-de Gennes (BdG) Hamiltonian, which in the thin-ring limit produces equations of motion for azimuthal phonons that are analogous to the Mukhanov-Sasaki equation that describes scalar curvature perturbations in cosmology.

The generation of entanglement constitutes the quantum signature of the pair-creation process and its observation would be a smoking gun for the  quantum origin of the observed density perturbations.  The goal of this paper is to quantify such entanglement and to propose ways of measuring it. Special attention has been paid in this article to include the effects of thermal noise, losses and detector inefficiencies. 

 Our analysis can be applied to any expansion history of the thin ring, as long as the ring is not expanding in the initial and final regions.  Having early- and late-time regions where
the radius of the ring is time independent, permits one to talk about particle creation, which in turn provides a way to talk about entanglement production. Note that in the real inflationary universe such ``in'' and ``out'' regions are not available, since the universe keeps expanding after inflation ends (and  what happened before inflation is not yet understood); this implies that there is no unambiguous way to define particle creation and, consequently, the quantification of entanglement becomes also ambiguous \cite{Agullo:2022ttg}. In this sense, being able to engineer non-expanding ``in'' and ``out'' regions is advantageous in the study of entanglement generation.

Techniques based on Gaussian states for continuous variable quantum systems are a powerful and efficient tool to quantify entanglement in this scenario. Our analysis has shown that, as one could intuitively expect, quantum entanglement is fragile to noise and losses. When losses can be  approximated  by a Gaussian channel, we have quantified the region in the parameter space where thermal noise and losses completely decohere the phonon pairs that would be otherwise entangled. Under such circumstances, all quantum signatures of the pair-creation process are gone, and one is left with the amplification of  thermal noise by the expanding ring, a process that can be entirely accounted for  in classical terms. Our analysis, therefore, helps to delineate the boundary where one could hope observing genuine quantum effects. 

We have further adventured and introduced ideas, adapted from \cite{Agullo:2021vwj,Brady:2022ffk}, to amplify the generation of entanglement, based on seeding the process with single-mode squeezed states, in order to compensate for the deleterious effects aforementioned, and to maintain the genuine quantum features generated by the expansion present in the final state. In addition, we have sketched a protocol (inspired
by the experiments of Hellweg et al.~\cite{Hellweg:2003}) to measure entanglement in the inflationary toroidal BEC.  Although our protocol
 may require a prohibitively large amount of practical resources, 
 it constitutes a concrete example of what would be needed to fully reconstruct the final phonon state and to quantify entanglement.

We conclude this section by suggesting some directions for future
work. We note that, although we have focused on an expanding BEC ring, our idea applies to other scenarios for simulating  pair-creation in expanding (or contracting) backgrounds, such as the two-dimensional quantum
field simulator considered in Refs.~\cite{Tolosa-Simeon:2022umw} and \cite{Viermann:2022wgw}.
 
Future studies could look into constructing protocols to detect entanglement in quantum states that have non-Gaussianity present in them. It would also be interesting to compare the procedure and results of this protocol with the experiments of Chen et al. \cite{Chen:2021xhd}, that explore how to witness entanglement inside a BEC using the Peres-Horodecki criterion. 

In this work, we studied how initial squeezing of one mode can amplify
entanglement~\cite{Agullo:2021vwj,Brady:2022ffk}; extensions of this
analysis can include the case where modes $n$ and $-n$ are individually
squeezed, which may provide even more amplification of entanglement 
in the final state.  Additionally, future work may investigate how such squeezing
of the initial states may be accomplished in a realistic toroidal BEC
experiment.

\section{Acknowledgements}
The authors are grateful to 
Anthony Brady, Lior Cohen, Adri\`a Delhom,  Stav Haldar, and 
Marlan Scully for useful comments and discussions. IA and DK are especially thankful to Anthony Brady; some of the tools used in this paper were developed in collaboration with him, and applied to analog Hawking radiation in optical systems. AB acknowledges financial support from the Department of Physics and Astronomy at LSU. 
AB and DES acknowledge financial support from NSF grant 
PHY-2208036.
IA and DK acknowledge financial support from the NSF grant PHY-2110273, and from the Hearne Institute for Theoretical Physics. 
JW acknowledges financial support from NSF grant DMR-2238895.
IA is also supported by the RCS program of  Louisiana Board of Regents through the grant LEQSF(2023-25)-RD-A-04, and 
in part by Perimeter Institute for Theoretical Physics. Research at Perimeter Institute is supported by the Government of Canada through the Department of Innovation, Science, and Economic Development, and by the Province of Ontario through the Ministry of Colleges and Universities. 
JW performed part of this work  at the Aspen Center for Physics, which is supported by National Science Foundation grant PHY-2210452.

\bibliography{Bibliography.bib}

\appendix

\section{Gaussian states and linear evolution: a short review} \label{appendix: Gaussian states}
\label{AppendixA}
For self-consistency, in this Appendix we provide a  summary of some elements of Gaussian states for continuous variable quantum systems and their evolution under quadratic Hamiltonians. See Ref.\onlinecite{Serafini:2017} for proofs omitted here. 
Note that we choose units in which $\hbar =1$ in this Appendix.  

\subsection{Evolution of linear systems with $N$ degrees of freedom}

Consider a dynamical system containing $N$ classical degrees of freedom. Quantum mechanically the system is described by $N$ pairs of canonically conjugate operators $\hat x_I,\hat p_I$, with  $I=1,\cdots,N$. Let us define a vector $\hat {\bf r}$ made of all canonical pairs: 
\be \hat {\bf r}=(\hat x_1,\hat p_1, \cdots, \hat x_N,\hat p_N)^{\top}\, .\ee
 In the following, we will assume that $\hat x_I$ and $\hat p_I$ have been re-scaled using the dimensionful constants of the problem under consideration to have  dimensions of action, as commonly  done when working with harmonic oscillators.
Let $\hat {\bf r}^i$ be the $i$-th component of $\hat {\bf r}$, where low case indices $i, j,\cdots$ run  from 1 to $2N$ (capital letter indices $I,J,\cdots$, instead, run from 1 to $N$). Then, the canonical commutation relation can be succinctly written as 
\be 
 [\hat{\bf r}^i,\hat{\bf r}^j]=i\, {\bf \Omega}^{ij},\hspace{0.7cm} {\bf \Omega}\equiv
 \bigoplus_N \begin{pmatrix}
 0 & 1\\
 -1 & 0
  \end{pmatrix} ,
\ee
where the anti-symmetric matrix ${\bf \Omega}$ is the (inverse of) symplectic form of the classical phase space. 

We are interested in processes where there exist  asymptotic regions in the past and future,  where one can define preferred canonical operators $\hat {\bf r}_{\rm (in)}$ and $\hat {\bf r}_{\rm (out)}$,  and an interaction region in between. We want to describe the scattering process between $\hat {\bf r}_{\rm (in)}$ and $\hat {\bf r}_{\rm (out)}$. In this article we restrict attention to systems  whose  Hamiltonians are  quadratic in the canonical variables $\hat {\bf r}$. Hence, the equations of motion are linear and, consequently,  the in and out modes are related  by a simple matrix multiplication:
\be \label{Sr} \hat{\bf r}_{\rm (out)}={\bf S} \cdot \hat{\bf r}_{\rm (in)}\, ,\ee
where  ${\bf S}$ is the scattering matrix. As shown below, this matrix can be obtained by solving the classical equations of motion (or equivalently, it can be constructed from the Bogoluibov coefficients). Since the time evolution of a closed system is a canonical transformation, ${\bf S}$ must leave the symplectic form invariant, ${\bf S}\cdot {\bf \Omega}\cdot {\bf S}^{\top}={\bf \Omega}$. In other words, 
 ${\bf S}$ must belong to the symplectic group, ${\bf S}\in {\rm Sp}(\mathbb{R}, 2N)$.

The scattering process can be equivalently formulated using annihilation and creation variables, instead of canonical operators. Define, for each canonical pair $(\hat x_I,\hat p_I)$, the non-Hermitian operator $\hat a_I=\frac{1}{\sqrt{2}} (\hat x_I +i\hat p_I)$. If we define the vector 
\be \hat{{\bf A}}\equiv (\hat{a}_{1},\hat{a}^{\dagger}_1,\cdots, \hat{a}_{N},\hat{a}^{\dagger}_N)\, , \ee
then the relation between $\hat{\bf r}$ and $\hat{{\bf A}}$ reads 
\begin{equation}\label{B}
\hat{{\bf A}}= {\bf B}\cdot \hat{\bf r},~~~~{\bf B}\equiv\bigoplus_N \frac{1}{\sqrt{2}}    \begin{pmatrix}
 1 & i\\
 1 & -i
\end{pmatrix} , \end{equation}
where we have denoted by ${\bf B}$ the ``change of basis'' matrix between $\hat {\bf A}$ and $\hat {\bf r}$.  In these variables,  the canonical commutation relations read 
\bea  [\hat{{\bf A}}, \hat{{\bf A}}]&=&{\bf B}\,  [\hat{{\bf r}}, \hat{{\bf r}}]\,  {\bf B}^{-1}\nonumber \\ &=& {\bf B}\, i\, {\bf \Omega}\,  {\bf B}^{\top}={\bf \Omega}\, , \nonumber \eea
This expression compactly captures the familiar commutation relation of annihilation and creation operators. 
 
The scattering matrix between in and out modes can now be written as
 \be \hat{\bf A}_{\rm (out)}={\bf S}_{(\bf A)} \cdot \hat{\bf A}_{\rm (in)}\, ,\ee
where ${\bf S}_{(\bf A)}$ is related to  ${\bf S}_{(\bf r)}$  by  ${\bf S}_{(\bf A)}= {\bf B}\cdot {\bf S}_{(\bf r)}\cdot  {\bf B}^{-1}$. 
It is common to refer to the components of  ${\bf S}_{(\bf A)}$ as Bogoluibov coefficients $\alpha_{IJ}$ and $\beta_{IJ}$ 
\begin{equation}\label{Bogoliubov Transformation Matrix}
{\bf S}_{(\bf A)}\equiv\begin{bmatrix}
	\alpha_{11} & \beta_{11} & \cdots  & \alpha_{1N} & \beta_{1N} &\\
	\beta^*_{11} & \alpha^*_{11} & \cdots  & \beta_{1N}^* & \alpha^*_{1N} &\\
	\vdots  & \vdots &\vdots  & \vdots & \vdots \\
	\alpha_{N1} & \beta_{N1} & \cdots  & \alpha_{NN} & \beta_{NN} &\\
	\beta^*_{N1} & \alpha^*_{N1} & \cdots  & \beta_{NN}^* & \alpha^*_{NN} &\\
\end{bmatrix}.
\end{equation}
The matrix ${\bf S}_{\bf (A)}$ also belongs to the symplectic group. The property ${\bf S}_{(\bf A)}\cdot  {\bf \Omega}\cdot  {\bf S}_{(\bf A)}^{\top}={\bf \Omega}$ is equivalent to the perhaps more  familiar constraints satisfied by  Bogoluibov coefficients: 
\bea \sum_K \Big(\alpha_{IK}\alpha^*_{JK}-\beta_{IK}\beta^*_{JK}\Big)
&=&\delta_{IJ} \nonumber \, ,\\  
\sum_K\Big(\alpha_{IK}\beta_{JK}-\alpha_{IK}\beta_{JK}\Big)&=&0 \, . \eea

\subsection{Gaussian states} We restrict our analyses to Gaussian states. Recall that Gaussian states $\hat \rho$, pure or mixed, are quantum states for which the quantum moments $\langle \hat{\bf r}^{i_1}\cdots \hat{\bf r}^{i_n}\rangle$ satisfy the same relations as the statistical moments of a Gaussian multi-variable probability distribution. This implies, in particular, that  the first and second moments completely determine the rest. Therefore, rather than working with the density matrix $\hat \rho$, which is infinite dimensional, one can alternatively  describe in full a Gaussian state by the $2N$ dimensional vector of its first moments 
\be
\label{Eq:mudef}
{\bf \mu}= \langle \hat{\bf r}\rangle,
\ee
and its symmetrized second moments (the so-called covariance matrix) 
\be\label{Eq:sigmadef}
{\bm \sigma}=\langle \{ (\hat{\bf r}-{\bm \mu}),(\hat{\bf r}-{\bm \mu}) \rangle, 
\ee
where the curly brackets denote the anti-commutator.\footnote{One subtracts ${\bm \mu}$ in the definition of ${\bm \sigma}$  to avoid having redundant information in the first and second moments. One focuses on the symmetric part of the second moments because the anti-symmetric part is  determined by the canonical commutation relations,  and is state independent. Therefore, the pair $(\bm{\mu},\bm{\sigma})$ is the minimum information needed to completely and uniquely characterize a Gaussian state.}
 
Gaussian states include vacua, coherent, thermal, and squeezed states. Therefore, although our analysis is restricted, the family of Gaussian states is sufficiently general to describe most of the states one can easily create and manipulate in the laboratory. To give a few examples, the vacuum of a set of $N$ oscillators is characterized by $\bm{\mu}=\bm{0},\,  \bm{\sigma}= \bm{\bm I}_{2N}$; a coherent state by $\bm{\mu}\neq \bm{0},\,  \bm{\sigma}= {\bm{I}}_{2N}$ and a 
thermal state $\bm{\mu}=\bm{0},\,  \bm{\sigma}= \bigoplus_I (1+2\, n_I)\, {\bm{I}}_{2}$, where $n_I$ is the mean number of thermal quanta in the mode $I=1,\cdots, N$. Thermal states are mixed Gaussian states. 
 
Many  properties of a Gaussian state can be extracted easily from $\bm{\sigma}$. One such propertiy that we use in the main body of this paper is the purity 
$P$. It is obtained from the covariance matrix of  the Gaussian state by $P(\sigma)=1/\sqrt{{\rm{det}\sigma}}$; it is one for pure states and smaller than one for mixed states. Note the purity does not depend on the first moments.

Evolving Gaussian states under quadratic Hamiltonians is very simple. The linearity of the evolution guarantees that  an initial Gaussian state $(\bm{\mu}^{\rm (in)},\bm{\sigma}^{\rm (in)})$ evolves to another Gaussian state $(\bm{\mu}^{\rm (out)},\bm{\sigma}^{\rm (out)})$ determined by
 
\begin{align} 
  \bm{\mu}^{\rm(out)}&=\bm{S}\bm{\mu}^{\rm(in)},\label{eq:out_mu_general}\\
  \bm{\sigma}^{\rm(out)}&=\bm{S}\bm{\sigma}^{\rm(in)}\bm{S}^\top.\label{eq:out_sigma_general}
\end{align}
 
\subsection{Entanglement in Gaussian states}  

Consider a partition of the system of $N$ modes in two subsystems,  each  made of a subset of the canonical pairs $(\hat x_I,\hat p_I)$ (these are Gaussian subsystems). The first moments and covariance matrix of a Gaussian state for the entire system have the following form 
\bea
{\bm \mu}_{AB}&=&({\bm \mu}_A,{\bm \mu}_{B})^{\top}\, , \\
{\bm \sigma}_{AB}&=&\begin{bmatrix} {\bm\sigma}^{\rm (red)}_{A} & {\bf C}_{AB}\\{\bf C}_{AB}^{\top} & {\bm\sigma}^{\rm (red)}_B  \end{bmatrix}  
\eea
where $({\bm \mu}_A,{\bm\sigma}^{\rm (red)}_{A})$ and $({\bm \mu}_B,{\bm\sigma}^{\rm (red)}_{B})$ describe the reduced Gaussian state of each subsystem individually. The matrix ${\bf C}_{AB}$ describes the correlations between the two subsystems; these correlations could be  classical  or contain entanglement. 

If the total state is pure, the von Neumann entropy of the reduced state of each of the subsystems are equal to each other, and it provides a faithful quantifier entanglement between $A$ and $B$---the so-called entanglement entropy. The von Neumann entropy of a  Gaussian state   $({\bm \mu},{\bm\sigma})$ for an $N$-mode system can be easily computed from the $N$  symplectic eigenvalues of  ${\bm\sigma}$, denoted by $\nu_I$, with $I=1\cdots N$. The symplectic eigenvalues  are equal to the modulus of the  eigenvalues of   the matrix $\bm{\sigma}^{ik}\bm{\Omega^{-1}}_{kj}$. The von Neuman entropy reads
\bea S[\bm\sigma]=\sum_I^N&\Big[& \left( \frac{\nu_I+1}{2}\right) \log_2\left( \frac{\nu_I+1}{2}\right)\\
&-&\left( \frac{\nu_I-1}{2}\right) \log_2\left( \frac{\nu_I-1}{2}\right)\Big] . \eea
If the total state is mixed, the von Newman entropy of the subsystems is no longer an entanglement measure. A convenient measure for pure and mixed states alike is the Logarithmic Negativity, $E_{\mathcal{N}}$.  The logarithmic negativity is in one-to-one correspondence with the violation of the  PPT criterion (Positivity of Partial Transpose) for quantum states \cite{Peres:1996dw,Plenio:2005cwa,Simon:1999lfr}, a criterion that all separable quantum states obey. For a Gaussian state made of two Gaussian sub-systems A and B, it is given by 
\be \label{logneg} E_{\mathcal{N}}[{\bm\sigma}]=\sum_I {\rm Max}[0, -\log_2 \tilde \nu_I]\, , \ee
where $\tilde \nu_I$ are the symplectic eigenvalues of the {\em partially transposed} covariance matrix $\bm{\tilde \sigma}$, defined from $\bm \sigma$ by reversing the sign of all components involving one momenta $\hat p_I$ of the subsystem B. If either of the Gaussian subsystems is made of a single mode, ($N_A=1$), regardless of the size of the other subsystem,  $E_{\mathcal{N}}$ is a faithful entanglement quantifier, in the sense that $E_{\mathcal{N}}=0$ if and only if the state is separable. It is also an entanglement monotone, and hence it can be used to quantify entanglement (see \cite{Serafini:2017} for further details).  For Gaussian quantum states, the value of the $E_{\mathcal{N}}$ has an operational meaning as the exact cost (measured in ``Bell pairs'' or entangled bits, ebits) that is required to prepare or simulate the quantum state under consideration \cite{Wilde:2020,alpha-Wilde:2020}.

In the analog gravity literature, there has been  focus on a particular Cauchy-Schwarz inequality to evaluate entanglement between two single-mode systems in a state $\hat \rho_{AB}$, first introduced in Refs.~\cite{Nova:2014} and further discussed in \cite{Busch:2013gna,Busch:2014bza}.
Consider the quantity 
\be
\label{eq:deltaDef}
\Delta\equiv \langle\hat n_A\rangle\,  \langle  \hat n_B \rangle  - |\langle \hat{a}_A\hat{a}_B\rangle|^2,
\ee
 where $\hat{a}_A$ and $\hat{a}_B$ are annihilation operators for each mode, and $\hat  n_A$ and $\hat  n_B$ are number operators defined from them, respectively. 
The inequality $\Delta<0$ is a {\em sufficient} condition for entanglement. It is not necessary though, in the sense that some entangled states do not violate the inequality. It is not an entanglement monotone either \cite{Brady:2022ffk}, even when restricted to Gaussian states. But it is a useful criteria to signal the presence of entanglement in many circumstances, particularly convenient because its evaluation requires only knowledge of  three moments $\langle\hat  n_A\rangle$, $\langle  \hat n_B \rangle$ and $\langle \hat{a}_A\hat{a}_B\rangle$. In contrast, $E_{\mathcal{N}}$ requires  knowledge of the entire covariance matrix, something that demands full state tomography on the two-mode system.

\section{The case of a $C^{2}$ scale factor} 
\label{appendix: C2 scale factor}
In this Appendix, we describe the expanding toroidal BEC dynamics with a  $C^{2}$ scale factor 
which describes asymptotically static regions in the past and future with  an inflationary phase in between. I.e., 
we study $a(t)$ that is continuous and also has continuous first and second time derivatives. This was studied analytically by Glenz and Parker~\cite{Glenz:2009zn} for the case of a scalar field in a $(3+1)$-dimensional spacetime.   

As discussed in the main text, scalar modes $\chi_{k}$ in the expanding toroidal BEC 
evolve on an analog FLRW background characterized by scale factor $a(t)$ and satisfy the Mukhanov-Sasaki equation~\cite{Sasaki:1983kd,Kodama:1985bj,Mukhanov:1988jd}:
\be\label{Mukhanov Sasaki FLRW}
\ddot{\chi}_{k} + \big(1+\gamma\big)\frac{\dot{a}}{a}\dot{\chi}_{k} +  \frac{k^{2}}{a^{2}}\chi_{k} = 0,  \\
\ee
which can be derived from Eq.~(\ref{MukhanovSasakiQP}) by setting $\alpha=1$, substituting $R(t) = R_0 a(t)$, and defining the wavenumber $k=nc/R_{0}$ in terms of the mode index. The entire evolution of the scale factor consists of three regimes:
\begin{equation}
a(t) = \begin{cases} a_i(t), & \text{for $t<t_1$},
\cr
a_{\rm inf}(t), & \text{for $t_1\le t\le t_2$},
\cr
 a_f(t), & \text{for $t>t_2$},
 \end{cases}
 \label{threeregimes}
\end{equation}
where $a_{\rm inf}(t)$ is an exponential function in proper time and the functions $a_i(t)$ and $a_f(t)$ are asymptotically constant in the future and past, and match (in a $C^2$ manner) with $a_{\rm inf}(t)$ at times $t_1$ and $t_2$, 
respectively.  To write explicit forms for these functions, it is convenient to switch to the harmonic time variable $\tau$, defined in terms of proper time $t$ by $d\tau=a(t)^{-(1+\gamma)}dt$. We emphasize that this is distinct from the parameter $\tau$ used in the main text to define the Hubble time scale.

The initial regime (denoted by subscript `$i$') has the following smooth scale factor:
\begin{equation}\label{Phase 1 scale factor}
a_{i}(\tau) = \Big(a_{1i}^{2\gamma}+(a_{2i}^{2\gamma}-a_{1i}^{2\gamma})n_{\text{F}}(-\tau/s_{i})\Big)^{\frac{1}{2\gamma}},
\end{equation}
where $n_{\text{F}}(x)=(e^{x}+1)^{-1}$ is the Fermi-Dirac function. The above scale factor approaches 
$a_{1i}$ at early times ($\tau\to-\infty$)  and approaches $a_{2i}$ at late times 
($\tau\to\infty$), with $s_{i}$ the timescale associated with this initial regime.  
Then, at some time $t_{1}$, an inflationary regime sets in where the scale factor is given by an exponential in proper time:
\begin{equation}\label{Phase 2 scale factor}
a_{\text{inf}}(t) = a(t_{1})e^{H_{\text{inf}}(t-t_{1})},
\end{equation}
where $H_{\text{inf}}$ is the constant Hubble parameter whose inverse determines the timescale of the 
expansion. To achieve $C^{2}$, we demand that at $t=t_{1}$, the scale factors $a_{i}$ and $a_{\text{inf}}$ are equal, and that the maximum value of the Hubble parameter $H(t)=a^{-1} da/dt$ of the first regime equals $H_{\text{inf}}$
(ensuring continuity of the first and second derivatives).  This helps us obtain the time $\tau_{i}=\tau_{i}(t_{1})$ at which the first regime smoothly joins onto the inflationary regime:
\begin{equation}
\label{tau-i}
\tau_{i} = s_{i}\log\frac{(1+\gamma)(a_{1i}^{2\gamma}-a_{2i}^{2\gamma})+C_{i}}{4\gamma a_{2i}^{2\gamma}}.
\end{equation}
From this, we obtain the scale factor $a(t_{1})$ at $t_{1}$ (see Eq.~(\ref{Phase 2 scale factor})):
\begin{equation}\label{a of tau-i}
a(\tau_{i}) = a(t_{1}) = \bigg[\frac{-(1+\gamma)(a_{1i}^{2\gamma}+a_{2i}^{2\gamma})+C_{i}}{2(-1+\gamma)}\Bigg]^{\frac{1}{2\gamma}}.
\end{equation}
Lastly, by equating the Hubble rate $H(t)$ at the junction $t=t_{1}$, we find the Hubble parameter $H_{\text{inf}}$ during inflation to be:
\begin{eqnarray}\label{Hinf}
H_{\text{inf}} & = & \frac{\big(-2\gamma(a_{1i}^{2\gamma}+a_{2i}^{2\gamma})+C_{i}\big)}{(a_{1i}^{2\gamma}-a_{2i}^{2\gamma})s_{i}(-1+\gamma)(1+3\gamma)} \nonumber \\
&\times& \bigg[\frac{-(1+\gamma)(a_{1i}^{2\gamma}+a_{2i}^{2\gamma})+C_{i}}{2(-1+\gamma)}\Bigg]^{-\frac{(1+\gamma)}{2\gamma}}.
\end{eqnarray}
In all these expressions, i.e., equations (\ref{tau-i}), (\ref{a of tau-i}) and (\ref{Hinf}), we used the function $C_{i}$ which is defined as follows:
\begin{eqnarray}
&&C_{i} = \\
&& \sqrt{(1+\gamma)^{2}a_{1i}^{4\gamma}+(14\gamma^{2}-4\gamma-2)^{2}a_{1i}^{2\gamma}a_{2i}^{2\gamma}+(1+\gamma)^{2}a_{2i}^{4\gamma}}. \nonumber
\end{eqnarray}

The preceding steps ensure that the first two regimes of Eq.~(\ref{threeregimes}) match in a $C^2$ manner at 
$t_1$. Now, we repeat these steps at time $t=t_{2}$, where inflation ends and the final regime begins (denoted by subscript `$f$').  We take the final regime scale factor to have a form similar to 
Eq.~(\ref{Phase 1 scale factor}):
\begin{equation}\label{Phase 3 scale factor}
a_{f}(\tau') = \Big(a_{1f}^{2\gamma}+(a_{2f}^{2\gamma}-a_{1f}^{2\gamma})n_{\text{F}}(-\tau'/s_{f})\Big)^{\frac{1}{2\gamma}},
\end{equation}
but with the replacement $i\to f$, and using a different time variable $\tau'$ for this regime. Again demanding the continuity of the scale factors $a_{f}(\tau')$ and $a_{\text{inf}}(t)$ at $t=t_{2}$ (equivalent to
$\tau'=\tau'_f$), we obtain the timescale of the final regime as $s_{f}=s_{i}(i\to f)$, and the joining time to be $\tau'_{f}=\tau_{i}(i\to f)$. (Obtained by replacing $i\to f$ in Eq.~(\ref{tau-i}).)
It is convenient to express these results in terms of the number of e-foldings $N$ that is defined to be logarithm of the ratio of the final scale factor and initial scale factor. Thus we write the following:
\begin{equation}
a_{2i}=a_{1i}e^{N_{i}},~~~a_{2f}=
{\color{red} a_{2i}}\, e^{N_{\text{inf}}+N_{f}}.
\end{equation}
In addition to this, if we also assume that the initial scale factor in the remote past is unity, i.e., $a_{1i}=1$, then we end up with four independent variables: $N_{i}$, $s_{i}$, $N_{\text{inf}}$, and $N_{f}$ that 
characterize the $C^2$ scale factor (\ref{threeregimes}) for the inflationary toroidal BEC.  Our next task is to solve Eq.~(\ref{Mukhanov Sasaki FLRW})  in all three regimes, matching the solutions.

To do this, it is convenient to re-write the Mukhanov-Sasaki equation (\ref{Mukhanov Sasaki FLRW}) in terms of harmonic time as follows:
\begin{equation}\label{MSE Harmonic Time}
\chi''_{k} + k^{2}a^{2\gamma}\chi_{k} = 0,
\end{equation}
where the prime $'$ denotes differentiation with respect to the harmonic time $\tau$. For the initial regime (\ref{Phase 1 scale factor}), the general solution to (\ref{MSE Harmonic Time}) is a linear combination of hypergeometric functions ${}_{2}F_{1}(a,b;c;d)$:
\begin{eqnarray}\label{general solution hypergeometric}
&&\chi_{k}(\tau) \\
&& = \frac{\delta_{1}(k)}{\sqrt{2ka_{1i}^{\gamma}}}e^{-ia_{1i}^{\gamma}k\tau} {}_{2}F_{1}(-a_{i}+b_{i},-a_{i}-b_{i};1-2a_{i};-e^{\tau/s_{i}}) \nonumber \\
&& + \frac{\delta_{2}(k)}{\sqrt{2ka_{1i}^{\gamma}}}e^{-ia_{1i}^{\gamma}k\tau} {}_{2}F_{1}(a_{i}+b_{i},a_{i}-b_{i};1+2a_{i};-e^{\tau/s_{i}}), \nonumber
\end{eqnarray}
where $\delta_{1,2}(k)$ are coefficients that are fixed by imposing initial conditions on the modes, $a_{i}=ika_{1i}^{\gamma}s_{i}$, and $b_{i}=ika_{2i}^{\gamma}s_{i}$. Assuming that the modes are in a vacuum state at early times $\tau\to-\infty$, we can take them to consist of only positive-frequency plane wave solutions:
\begin{equation}
\lim_{\tau\to-\infty}\chi_{k}(\tau) = \frac{1}{\sqrt{2ka_{1i}^{\gamma}}}e^{-ia_{1i}^{\gamma}k\tau}.
\end{equation}
This condition helps us pick out the correct form of the mode functions in the initial regime to be:
\begin{eqnarray}
&&\chi_{k}(\tau) = \\
&& \frac{1}{\sqrt{2ka_{1i}^{\gamma}}}e^{-ia_{1i}^{\gamma}k\tau} {}_{2}F_{1}(-a_{i}+b_{i},-a_{i}-b_{i};1-2a_{i};-e^{\tau/s_{i}}). \nonumber
\end{eqnarray}
In the inflationary regime with the exponential scale factor (\ref{Phase 2 scale factor}), the Mukhanov-Sasaki equation yields the following mode solution:
\begin{eqnarray}
&&\chi_{k}(t) = \frac{i}{2}\sqrt{\frac{\pi}{H_{\text{inf}}}}a_{\text{inf}}^{-\frac{1+\gamma}{2}}(t) \times \nonumber \\
&&  \bigg[E(k)H^{(1)}_{\frac{1+\gamma}{2}}\Big(\frac{k}{a_{\text{inf}}(t)H_{\text{inf}}}\Big) - F(k)H^{(2)}_{\frac{1+\gamma}{2}}\Big(\frac{k}{a_{\text{inf}}(t)H_{\text{inf}}}\Big)\bigg] , \nonumber
\end{eqnarray}
where $H^{(1)}$ and $H^{(2)}$ are Hankel functions of first and second kind respectively, and the coefficients $E(k)$ and $F(k)$ are fixed by matching the modes and their first derivatives at $t=t_{1}$. 
Finally, the solution to Mukhanov-Sasaki equation in the final regime (\ref{Phase 3 scale factor}) is similar to (\ref{general solution hypergeometric}) but with $i\to f$:
\begin{eqnarray}
&&\chi_{k}(\tau') \\
  &&\!\!\!\!\!\!\!\!
  = \frac{C(k)}{\sqrt{2ka_{1f}^{\gamma}}}e^{-ia_{1f}^{\gamma}k\tau'} {}_{2}F_{1}(-a_{f}+b_{f},-a_{f}-b_{f};1-2a_{f};-e^{\tau'/s_{f}}) \nonumber \\
&& + \frac{D(k)}{\sqrt{2ka_{1f}^{\gamma}}}e^{ia_{1f}^{\gamma}k\tau'} {}_{2}F_{1}(a_{f}+b_{f},a_{f}-b_{f};1+2a_{f};-e^{\tau'/s_{f}}), \nonumber
\end{eqnarray}
where $a_{f}=ika_{1f}^{\gamma}s_{f}$, and $b_{f}=ika_{2f}^{\gamma}s_{f}$, and the coefficients $C(k)$ and $D(k)$ are fixed by matching the mode functions and their first time derivatives at $t=t_{2}$. At late times, the scale factor (\ref{Phase 3 scale factor}) approaches the constant value $a_{2f}$, and thus we expect that the modes behave as a linear combination of positive and negative frequency plane waves:
\begin{equation}
\lim_{\tau\to\infty}\chi_{k}(\tau') \sim \frac{1}{\sqrt{2ka_{2f}^{\gamma}}} \big(\alpha_{k}e^{-ia_{2f}^{\gamma}k\tau'} + \beta_{k}e^{ia_{2f}^{\gamma}k\tau'}\big),
\end{equation}
where $\alpha_{k}$ and $\beta_{k}$ are the Bogoliubov coefficients that satisfy $|\alpha_{k}|^{2}-|\beta_{k}|^{2}=1$. Following the steps in Ref.~\cite{Glenz:2009zn}, we get the following expressions for the Bogoliubov coefficients: 
\begin{eqnarray}
&&\alpha_{k} = \bigg(\frac{a_{2f}}{a_{1f}}\bigg)^{\gamma/2} \Big(C(k)B(k)+D(k)B_{t}(k)\Big),~~~~~ \\
&&\beta_{k} = \bigg(\frac{a_{2f}}{a_{1f}}\bigg)^{\gamma/2} \Big(C(k)A(k)+D(k)A_{t}(k)\Big),~~~~~
\end{eqnarray}
where we define the following functions:
\begin{eqnarray}
A(k) & = & \frac{\Gamma(1-2a_{f})\Gamma(2b_{f})}{\Gamma(-a_{f}+b_{f})\Gamma(1-a_{f}+b_{f})}, \nonumber \\
B(k) & = & \frac{\Gamma(1-2a_{f})\Gamma(-2b_{f})}{\Gamma(-a_{f}-b_{f})\Gamma(1-a_{f}-b_{f})},
\end{eqnarray}
and the other two functions are related to these via $A_{t}(k)=A(k)[a_{f}\to -a_{f}]$ and $B_{t}(k)=A(k)[b_{f}\to -b_{f}]$. 

\begin{figure}[h!]
 \vspace{0.5cm}
	\begin{center}
	\includegraphics[width=1.0\columnwidth]{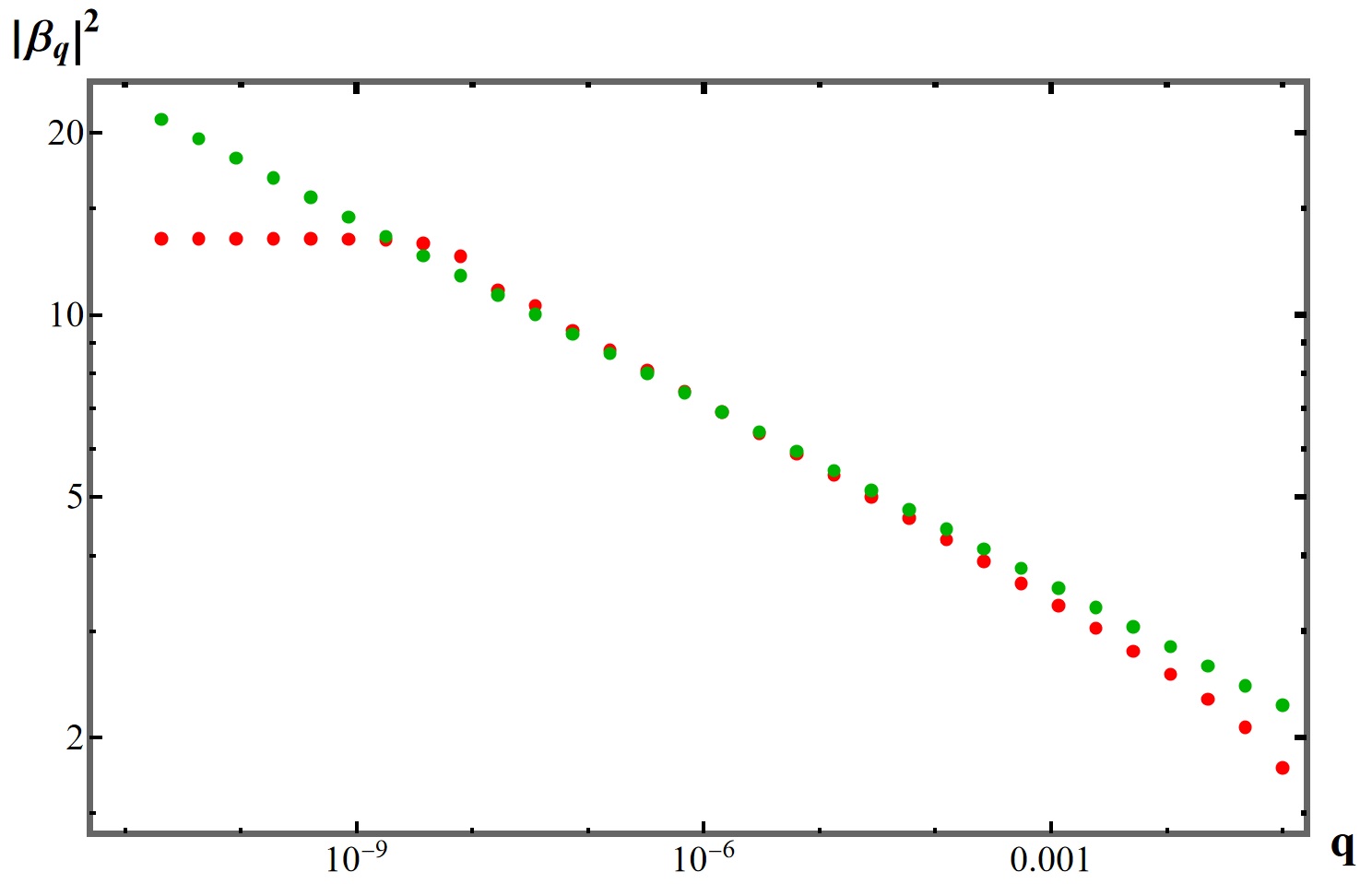}
	\end{center}
	\caption{Log-log plot of particle creation number $|\beta_{q}|^{2}$ versus the normalized wave number $q$ (red circles). For small wave numbers, $|\beta_{q}|^{2}$ is constant, for intermediate wave numbers it shows a universal linear behavior due to inflation, and for large wave numbers it rapidly decays. In the linear inflationary regime, $|\beta_{q}|^{2}\sim q^{-\gamma}$ and thus the slope of the log-log plot is $\gamma$ (green circles, which decay exactly as $q^{-\gamma}$, have been added for comparison). For this plot, and without loss of generality, we take in the initial regime the scale factor to be $a_{1i}=1$, number of e-folds $N_{i}=\log(1.1)$ and timescale for expansion $s_{i}=1$. We chose the inflationary regime to have a large number of e-folds $N_{\text{inf}}=20$, whereas for the final regime we took $N_{f}=\log(1.1)$, and we chose the quantum pressure to be $\gamma=0.1$.}
	\label{beta}
\end{figure}

The preceding equations determine the particle creation number $|\beta_k|^2$ for an inflationary toroidal BEC undergoing
an expansion that is $C^2$ everywhere with a exponential (``de Sitter''-like) central regime.  Next, we show that this particle creation
exhibits a universal behavior reflecting intrinsic properties of the inflation (such as the damping parameter 
$\gamma$).  To do this, in 
Fig.~\ref{beta}, we show a log-log plot of the particle creation probability $|\beta_{k}|^{2}$ versus the normalized wave number $q$ defined as follows:
\begin{equation}
q = \frac{k}{\sqrt{\frac{\gamma}{2}+\frac{\gamma^{2}}{4}}H_{\text{inf}}a_{f}(\tau'_{f})}.
\end{equation}
To focus on inflationary physics, we chose the number of efolds to be small in the initial and final regimes ($N_{i}=N_f = \log(1.1)$) and large in the de Sitter regime ($N_{\rm inf} = 20$).  Fig.~\ref{beta} shows that there exists an intermediate regime of wave numbers for which the modes only experience the inflationary expansion with the particle creation spectrum showing the power-law behavior:
\begin{equation}
|\beta_{q}|^{2} \sim q^{-\gamma},
\end{equation}
that reflects the damping parameter $\gamma$.  This is precisely what is expected from an analysis of the asymptotic behaviour of the Hankel function $H^{(1)}_{\frac{1+\gamma}{2}}$.
Thus, the particle-production spectrum in a expanding toroidal 
BEC indeed shows universal behavior
due to inflation when the scale factor is $C^2$~\cite{Glenz:2009zn}.

Note that in the main text we have focused on a $C^0$ expansion, i.e., the radius $R(t)$ (or the scale factor $a(t)$) is continuous, but its derivatives are not. In such cases, we get a different power-law $|\beta_{q}|^{2}\sim q^{-2}$ at asymptotically large mode wavevector (see Eq.~(\ref{Asymptotic Particle Creation}))
with the power-law $|\beta_{q}|^{2}\sim q^{-\gamma}$ holding at
intermediate wavevectors (see Eq.~(\ref{Asymptotic Particle Creation Superhorizon})).  In addition, the $C^0$ case exhibits
oscillations.   One might ask whether the $C^{2}$ case with insignificant initial and final regimes, (i.e., $N_{i}$, $s_{i}$, $N_{f}$ and $s_{f}$ are all very small) is able to reproduce these features of the $C^0$ case. We find that in this limit, $|\beta_{q}|^{2}$ pertaining to the $C^{2}$ case indeed starts exhibiting oscillations, but its overall magnitude and power-law with respect to the mode index do not agree with the $C^0$ case. Thus some of the features of the $C^0$ case are not rooted entirely in the inflationary regime, and some of them result just from the $C^0$ character of $R(t)$ (a discontinuity in the derivative), i.e., they would not appear for any smooth $R(t)$. In the lab, any $R(t)$ that one can create is smooth, and in this sense some of the features of the $C^0$ could be called mathematical artifacts. On the other hand, the $C^0$ expansion has advantages in that it allows us to solve for the $\beta$ coefficients analytically. 
Additionally, even though the $C^2$ case more plausibly avoids such
artifacts,  the universal linear feature due to inflation only appears in the limit of large number of inflationary e-foldings, which might not be possible to achieve in a real experiment.   

In any case, the formulas developed in this paper apply equally well to all expansion histories that are  time-independent in the past and future.

\end{document}